\documentclass[notitlepage,aps,prd,superscriptaddress,nofootinbib]{revtex4-1}

\usepackage[sc]{mathpazo} 
\usepackage[T1]{fontenc} 
\usepackage{microtype} 

\usepackage[hmarginratio=1:1,top=32mm,columnsep=20pt]{geometry} 
\usepackage[hang, small,labelfont=bf,up,textfont=it,up]{caption} 
\usepackage{booktabs} 
\usepackage{float} 
\usepackage{hyperref} 
\usepackage{graphics}
\usepackage{graphicx}
\usepackage{float}
\usepackage{amssymb}
\usepackage{slashed}
\usepackage{fancybox}
\usepackage{amsfonts}
\usepackage{epsfig}
\usepackage{rotating}
\usepackage{amsmath}
\usepackage{amsmath}%
\usepackage{wasysym}%
\usepackage{braket}
\usepackage{epsfig}
\usepackage{lettrine} 
\usepackage{paralist} 
\usepackage[utf8]{inputenc}

\usepackage{fancyhdr} 
\allowdisplaybreaks

\begin{document}


\title{Asymmetric Dark Matter and the hadronic spectra of hidden QCD}


\author{Stephen J. Lonsdale}
\email{lsj@student.unimelb.edu.au}
\affiliation{ARC Centre of Excellence for Particle Physics at the Terascale, School of Physics,\\ The University of Melbourne, Victoria 3010, Australia}
\author{Martine Schroor}
\affiliation{Van Swinderen Institute for Particle Physics and Gravity, University of Groningen, Nijenborgh 4, 9747 AG Groningen, The Netherlands}
\author{Raymond R. Volkas}
\affiliation{ARC Centre of Excellence for Particle Physics at the Terascale, School of Physics,\\ The University of Melbourne, Victoria 3010, Australia}




                              \renewcommand \thesection{\arabic{section}}
\renewcommand \thesubsection{\arabic{section}.\arabic{subsection}}

\begin{abstract}
The idea that dark matter may be a composite state of a hidden nonabelian gauge sector has received great attention in recent years. 
Frameworks such as asymmetric dark matter motivate the idea that dark matter may have similar mass to the proton, while 
mirror matter and $G \times G$ grand unified theories provide rationales for additional gauge sectors which may have minimal 
interactions with standard model particles. In this work we explore the hadronic spectra that these dark QCD models can allow. 
The effects of the number of light colored particles and the value of the confinement scale on the lightest stable 
state, the dark matter candidate, are examined in the hyperspherical constituent quark model for baryonic and mesonic states.
\end{abstract}
\maketitle 

\thispagestyle{fancy} 


\section{ \bf Introduction}\label{sec:Intro}
The current evidence of the makeup of the universe shows that one quarter of the energy budget is made up of dark matter\cite{Ade:2015xua}. The nature of dark matter is one of the key questions of present day physics.
The near equality of the baryon and dark matter mass densities,
\begin{equation}
\Omega_{DM} \simeq 5 \, \Omega_{B},
\end{equation}
motivates the idea that the origin of dark matter is connected to that of baryonic matter. If one expresses the critical density 
as a product of particle number density and mass, $\Omega_{DM} = n \, m$, then assuming a mechanism that generates similar number densities of matter and dark matter, we require the relation $ m_D \sim 5 \, m_p$. 
Asymmetric dark matter models provide a way to generate similar number 
densities \cite{Davoudiasl:2012uw, Petraki:2013wwa, Zurek:2013wia}. 
In these models one typically generates a dark matter asymmetry just like that of the visible sector
such that all dark antimatter and a large fraction of dark matter has been annihilated away into dark and/or standard radiation and
only dark matter remains. 
\\\\
In order to explain the similarity in mass we first note that the mass scale of ordinary baryons is generated by the confinement 
scale of QCD where the SU(3) gauge coupling becomes non-perturbative. 
If dark matter has a mass similar to the proton, then it is natural to consider dark confining gauge groups of a hidden sector. 
These 'dark sectors' have been considered in a large number of models in recent years and dark QCD models \cite{Higaki:2013vuv, Detmold:2014qqa, Bai:2013xga, Newstead:2014jva, Krnjaic:2014xza, Boddy:2014yra, Yamanaka:2014pva, Hardy:2014mqa, Antipin:2015xia} in particular have received 
much attention.\footnote{For other contexts in which dark matter bound states have been studied see, for example, Refs.\cite{MarchRussell:2008tu, Shepherd:2009sa, Asadi:2016ybp, Kouvaris:2016ltf, Laha:2013gva, Pospelov:2008jd, Kaplan:2009de, Petraki:2011mv, Cline:2012is, Kaplan:2011yj, Behbahani:2010xa, CyrRacine:2012fz, Laha:2013gva, Cline:2013pca, Petraki:2014uza, Wise:2014jva,Wise:2014ola, Petraki:2016cnz, Petraki:2015hla, Laha:2015yoa, Cirelli:2016rnw, Baldes:2017gzw, vonHarling:2014kha}.}
In these models almost all of the mass density of the universe would by dynamically generated.
The mass of a proton can by dimensional analysis alone be seen to depend largely on the confinement scale $\Lambda_{QCD} \sim 200 \, \text{MeV}$ and 
so as a first estimate we can relate the masses of dark baryons to a dark confinement scale, $\Lambda_{DM}$,  of a hidden SU(3) group.
The similarity in confinement scales at low energy can be explained by models such as mirror matter \cite{Foot:1991bp, Foot:2014mia} or $G \times G$ unification. 
Models where at high energy the gauge forces are described by a mirror symmetric $SU(5)_{DM} \times SU(5)_{VM}$, can feature spontaneously
broken mirror symmetry \cite{Lonsdale:2014yua, Lonsdale:2014wwa} resulting in two distinct sectors that have gauge couplings which run independently. 
We can then have the standard model consisting of the usual
$SU(3) \times SU (2) \times U (1)$, and a dark sector containing at least
$SU(3)_{D}$. 
\\\\
In this work we examine some of the properties that dark baryons of a hidden QCD may possess. 
We will consider the hidden sector colored fermions to be charged at most under a mirror copy of the SM gauge group $G' = SU(3)_{D} \times SU(2)_{D} \times U(1)_{D}$. 
In the context of broken mirror symmetries the similar confinement scales of the two SU(3) groups then has a natural explanation in the UV unification of the two non-abelian gauge couplings.
While exploring the rich spectra of such hidden theoretical QCD models may be 
beyond the reach of current dark matter experiments, the ground state and its dependence on a number of assumptions in the theory can be computed. 
For a model of dark matter we are primarily interested in dark QCD models and their lightest stable ground states, as on the timescale of the universe almost all dark matter would be expected to be in this form. 
Various theoretical models of QCD have seen 
great success in modeling the properties of baryons and mesons.
Lattice QCD, chiral perturbation theory, the relativistic and non-relativistic quark models, the bag model and others have all seen
encouraging results though no model other than lattice QCD can promise complete success.
The study of a dark QCD could also be approached in various ways. In this work we use a simple model of predicting the baryonic 
ground state spectra that captures the key dependence on confinement scale and has replicated the ground states of baryons well: the hyperspherical constituent quark model (hCQM) \cite{Richard:1992uk, Giannini:2012vy}. 
In section \ref{sec:HP} we review the 
hyperspherical constituent quark potential models and the specialization to hypercentral potentials
for the calculation of ground states. Section \ref{sec:HS} then explores the baryon and meson spectra and analyses a number of key scenarios.
We then consider the possible evidence for these cases and the cosmological consequences in Section \ref{sec:NDM} before concluding in Section \ref{sec:Conclusion}.

\section{Hyperspherical Potentials}\label{sec:HP}
To examine the masses of dark baryons we use the hyperspherical constituent quark model (hCQM) \cite{Richard:1992uk, Giannini:2012vy}. 
Potential models of the hadronic spectra have a long history in the development of the quark model. The hypercentral approximation is a special case of the 
hyperspherical formalism with a potential that depends on a single radius.  
The more accurate full hyperspherical formalism for three-body problems can be used, which reduces the problem to a set of infinite 
coupled differential equations in 6 dimensions \cite{Richard:1992uk}. However truncating the infinite set at 
first order, which is equivalent to the hypercentral approximation, already fits the experimental spectra quite well as noted in \cite{Richard:2012xw, Giannini:2003xx} and works particularly well on ground states,
which are the most important for the present work on examining dark matter candidates. 
First we rewrite the coordinates in the Jacobi form assuming the case of equal constituent quark masses, 
\begin{align}
\label{Jacobi coor}
\vec{\rho} &= \vec{r_2} - \vec{r_1} \nonumber \\ 
\vec{\lambda} &= \frac{1}{\sqrt{3}} (2\vec{r_3} - \vec{r_1} - \vec{r_2}) \\ 
\vec{R} &= \frac{1}{3}(\vec{r_1} + \vec{r_2} + \vec{r_3}). \nonumber
\end{align}
The center of mass motion is separated out, hence the coordinate associated with it, $R$, is not used. We then convert to hyperspherical polar coordinates,
\begin{align}
x &= \sqrt{\rho^2 + \lambda^2} \\
\phi &= \text{tan}^{-1}(\frac{\rho}{\lambda}) 
\end{align}
with $x$ the new hyperradius and $\phi$ the hyperangle. The 6D vector depends not only on $x, \, \phi$, but on the four angles associated with coordinates $\rho$ and $\lambda$ as well. 
These are denoted by two-vectors $\Omega_{\rho}, \, \Omega_{\lambda}$ respectively. 
In analogy with one-particle quantum mechanics we can introduce hyperspherical harmonics $Y_{[\gamma]}(\Omega_5)$, in analogy with the spherical harmonics in three dimensions,  and a six dimensional grand orbital 
angular momentum operator $L(\Omega_5)$, with $\Omega_5$ consisting of all five angles $\phi, \, \Omega_{\rho}, \, \Omega_{\lambda}$ 
and $l_{\rho,\lambda}$ are the conventional three dimensional orbital angular momenta quantum numbers corresponding to $\rho$ and $\lambda$, taking values (0,1,2...).  The eigenvalues of $L^2$ are given by
\begin{equation}
L^2 \, Y_{[\gamma]}(\Omega_5) = \gamma(\gamma + 4) \, Y_{[\gamma]}(\Omega_5).
\end{equation}
As the hypercentral formalism assumes that the potential is dependent only on the hyperradius we have that $V(\rho,\lambda)$ becomes $V(x)$
and the spatial wavefunction can be separated into functions of the hyperradius and the remaining angles,
\begin{equation}
\label{eq:hypercentralwavefunction}
\psi_{space} = \psi_{N,[\gamma]}(x) \, Y_{[\gamma]}(\Omega_5).
\end{equation}
The non-negative integer N now characterizes the radial excitations of the baryons and $\gamma$ is the grand angular quantum number such that $\gamma= 2n + l_\rho + l_\lambda$ with n a new quantum number, a non-negative integer (0,1,2...) 
that identifies the degree of the Jacobi polynomials that serve in the definition of the the hyperspherical harmonics.
$l_\rho$ and $l_\lambda$ are the angular momenta for the $\rho$ and $\lambda$ coordinates respectively.
For instance, the nucleon is associated to the $E_{N=0, \gamma=0}$ eigenvalue while its first radial excitation is $E_{10}$, otherwise known as the Roper resonance. 
Then the hyperradial Schr\"odinger equation becomes
\begin{equation}
\label{eq:hyperradial equation}
\left[\frac{d^2}{dx^2} + \frac{5}{x}\frac{d}{dx} - \frac{\gamma(\gamma+4)}{x^2}\right] \, \psi_{N,[\gamma]} = -2m\left[E_{N [\gamma]}-V(x)\right] \, \psi_{N,[\gamma]}
\end{equation}
which can be solved analytically for some select choices of $V(x)$ and otherwise will be solved for numerically.
In this work we use the matrix methods of \cite{Richard:1992uk} to solve for these eigenvalues.
One of the simplest quark interaction forms is described by the Cornell-type potential, 
\begin{equation}
\label{eq:cornell}
 V(x) = -\frac{\tau}{x} + k x,
\end{equation}
consisting of a hyperCoulombic and a linear term.
Other potentials with variations on this general shape have been considered in the literature of the constituent quark model \cite{Giannini:2012vy, Richard:1992uk, Isgur:1977ef, Isgur:1978xj}. These models have yielded
great successes in replicating the spectra when the parameters of the potential are fitted to just a few 
experimental states, though predicting the masses of both the light and heavy spectra simultaneously has presented more of a challenge. 
In the next section we examine the spectra for light baryons generated by such potentials and the implications for 
dark non-abelian composite states.

       \section{Hadronic Spectra}\label{sec:HS}
       While the constituent quark model has achieved considerable success in replicating the ground states of the baryon spectrum, 
       in adapting this method to the exploration of a hidden QCD we have to deal with the fact  
       that the parameters of the theory are taken from the experimental spectra, which are unavailable in the dark sector. 
       In the case of hyperspherical potential models we adopt a method of exploring the possible dark matter spectra by considering simplified 
       models that depend on only a few parameters, and then deducing how those parameters change for the dark QCD case. In particular we consider models in which the lightest dark quarks have near degenerate mass and where the baryon mass scale $E_0 \sim \Lambda_{DM}$ is a free parameter. 
       This is based on our treatment of $\Lambda_{DM}$ itself as a free parameter of the theory following the work in \cite{Lonsdale:2014wwa}. 
       In that model of asymmetric dark matter the UV value of $\alpha_{DM}$ is fixed by the $Z2$ symmetry imposed on the $G \times G_{mirror}$ theory. $\Lambda_{DM}$ will then vary from $\Lambda_{QCD}$ by the location of mass thresholds of 
       dark quarks with mass greater than $\Lambda_{DM}$, making the latter parameter different from the SM value, $\Lambda_{QCD} \sim 200 MeV$, following the mechanism of asymmetric symmetry breaking \cite{Lonsdale:2014wwa}. 
       In these models, the heavy dark quarks may have masses significantly larger than the dark confinement scale: $m_q >> \Lambda_{DM}$.
       As these very massive degrees of freedom have no effect, except through their production of a given $\Lambda_{DM}$, on the ground state of the dark SU(3) theory, the locations of the thresholds can be made to produce a given low scale value of $\Lambda_{DM}$ and thus $E_0$. 
       The lightest dark quarks with negligible bare mass then have a dressed mass of $\sim E_0/3$.
       \\\\
       With the variation of $\Lambda_{DM}$ we also vary accordingly 
       the length scale of the inter-quark potential and any other dimensional parameters associated with $\Lambda_{DM}$ in the theory. 
       As in the bag model of QCD, where the length scale of confinement scales inversely with the energy scale, we will vary the length scale 
       of the potential and constituent quark masses with $\Lambda_{DM}$. The parameters of the reference potential will be fitted to the standard model QCD spectra and a number of different potential forms will be used for comparison. 
       We define the ratio of confinement scales in the two sectors as 
       \begin{equation}
       \label{eq:ratiocon}
       \xi=\frac{\Lambda_{DM}}{\Lambda_{QCD}}.
       \end{equation} 
       In a model with high scale mirror symmetry, large values of $\xi$ are less likely since mass thresholds only vary the rate of running slightly and so similar 
       QCD mass scales for the two sectors are well motivated by the insensitivity of the scale of dimensional transmutation to higher mass scales in the theory \cite{Lonsdale:2014yua}.
       In such models, the Yukawa couplings of the two sectors are also independent despite the high scale mirror symmetry. This can be seen as an effect of both the different running couplings and the fact that
       the Higgs mechanisms responsible for mass generation in the two sectors can involve scalar states that are not mirror partners \cite{Lonsdale:2014wwa}. 
       In this work we similarly take the Yukawa coupling constants of the dark quarks to be effectively unrelated to those of the corresponding ordinary quarks.
       \\\\
       If $\xi>1$ and the lightest dark quark bare masses are comparable to the up and down quark of the SM, then the approximate chiral flavour symmetry becomes more exact as $\xi$ increases.
       This can be compared with our own QCD where it is the small  quark masses relative to the confinement scale that generates the isospin symmetry. 
       No symmetry connecting the bare up and down quark masses is necessary for strong isospin, only that they are small enough 
       to be insignificant compared to the near equal constituent masses the quarks gain from chiral symmetry breaking within the bound states in the constituent quark model. 
       It is in this sense that a dark QCD with $n_l$ light flavours with masses $m_l << \Lambda_{DM}$ can form an $SU(n_l)$ analogue of strong isospin. 
       The assumption of near degenerate dark quarks is then seen to refer to constituent masses. 
       This dark isospin is then a consequence of any hidden QCD where fermions gain mass via a Higgs-like mechanism and the product of the hidden sector Higgs VEV $v_D$ and any 
       dark quark Yukawa couplings are small in relation to the dark confinement scale.
       As the light dark quarks form the lightest bound states of a dark QCD, any colored fermions more massive than the confinement scale
       will  decay to lighter states of the theory and only be produced in small numbers following the dark quark hadron phase transition. 
       We also consider states analogous to the strange quark of QCD which have a mass which can be less than $\Lambda_{QCD}$ but can still contribute a significant amount
       to hadronic masses. There also exists the limiting case of when all of the dark quarks have mass above $\Lambda_{DM}$, which has been explored in models such as \cite{Forestell:2016qhc, Soni:2016gzf, Boddy:2014qxa}
       where the possibility of glueball dark matter is examined. In the case of six massless quarks  and a baryonic mass scale 
       $E_0 \sim \Lambda_{DM}$, \footnote{The effects of the strange quark as a virtual state contributing to the mass of the proton has a long history. 
       It is estimated that a massless strange quark may lower the nucleon mass scale of QCD by between $ \sim 1-20 \% $ \cite{Scadron:2006yw}.} the meson states will have zero mass as genuine Goldstone bosons. 
       The exception is the Goldstone Boson associated with the breaking of the anomalous U(1) axial symmetry. In QCD it is the $\eta'$. 
       Importantly, in one flavour QCD, the only meson of the theory will gain an anomalous mass and so even in a dark QCD model 
       with a single massless quark, there will be no massless Goldstone bosons. We will briefly discuss this particular case in the next section.\\
       
\begin{figure}[!ht]
      \begin{minipage}{\textwidth}
       \centering
       \includegraphics[width=.48\textwidth]{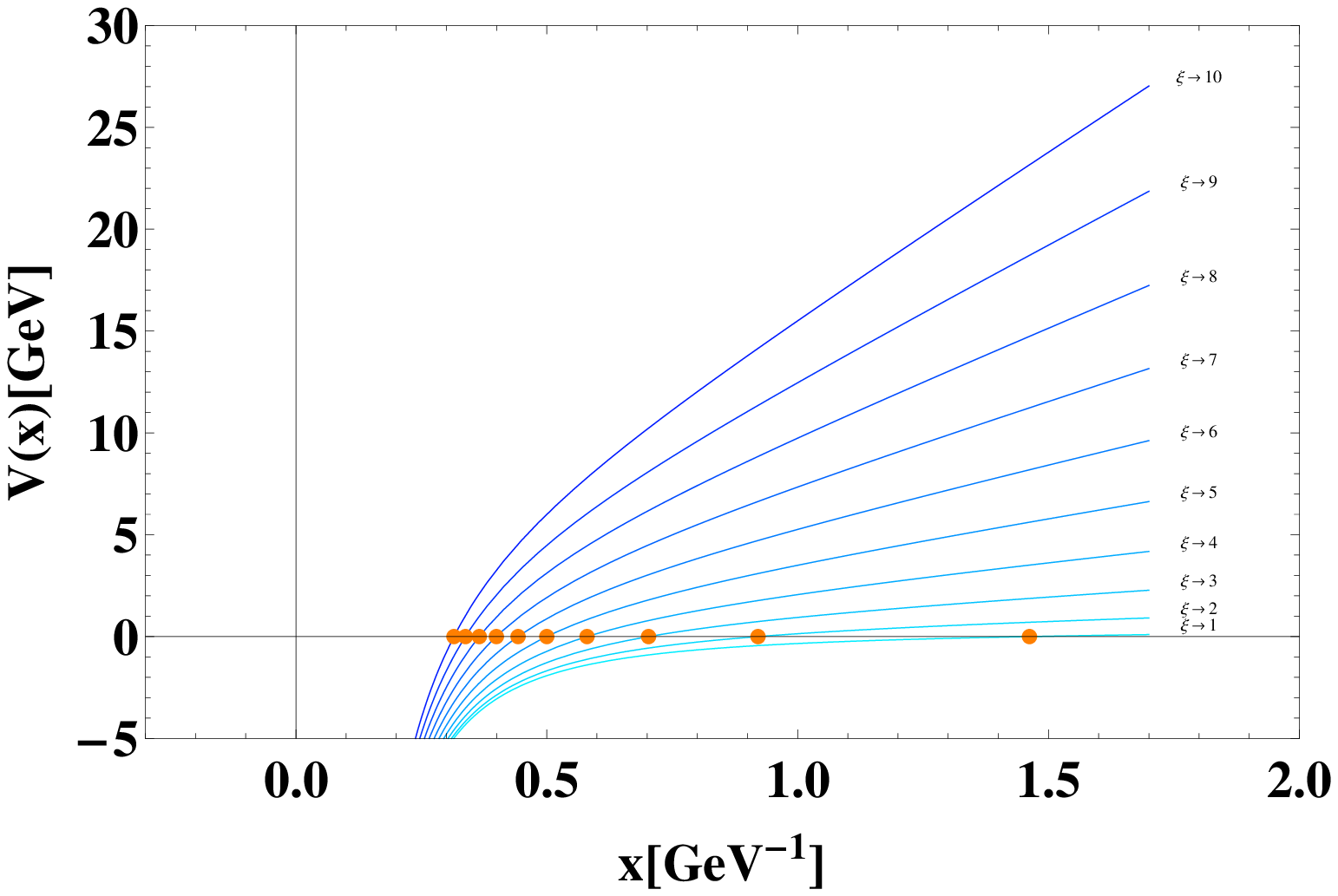}\quad %
       \includegraphics[width=.48\textwidth]{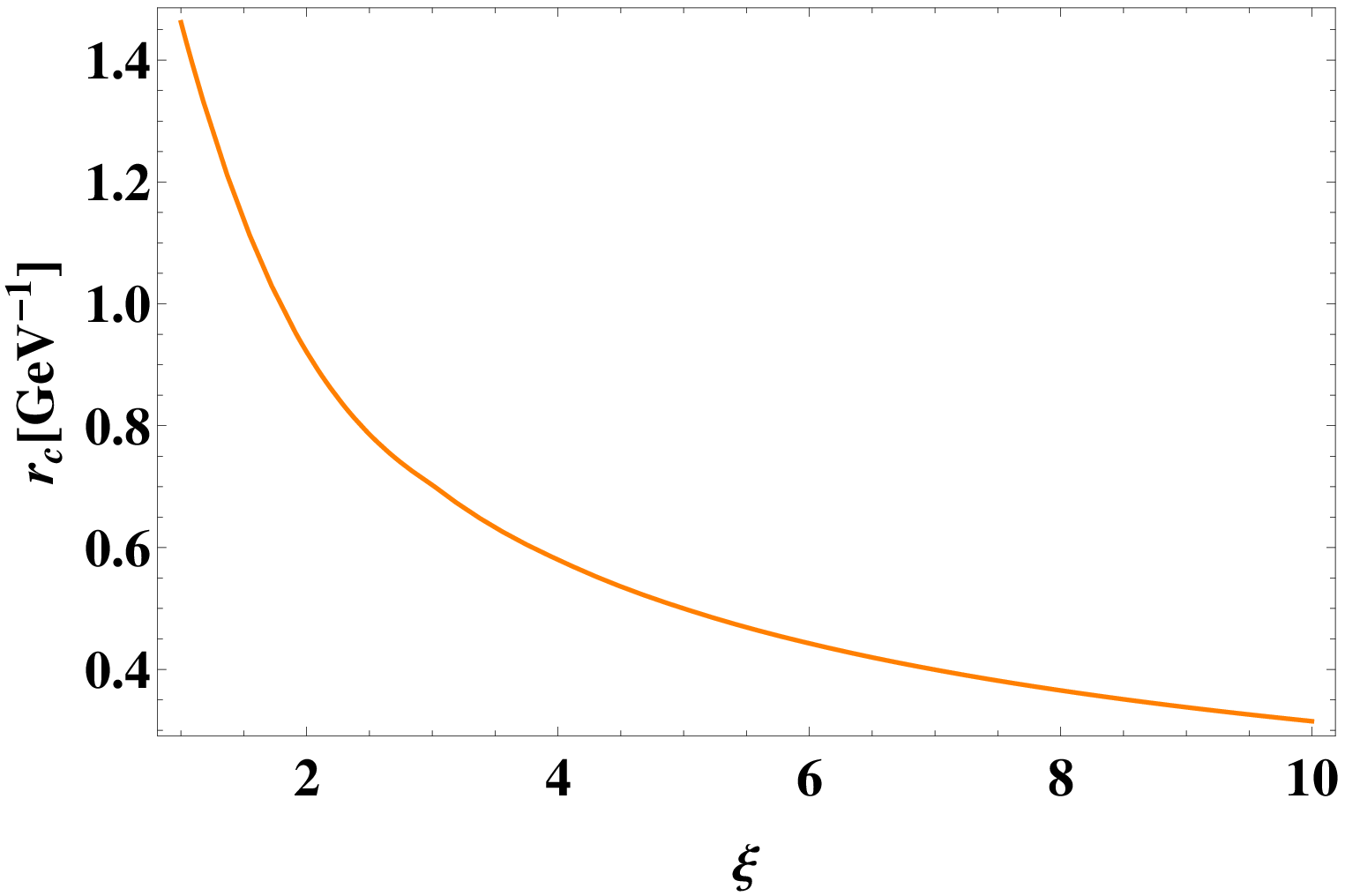}\\   %
       \caption{The variation of an example hypercentral potential V(x) with $\xi$ (left) and the value of $r_c$, an estimate of the radius of confinement, as a function of $\xi$ (right). 
       By scaling the dimensionful parameters of the interaction potential with confinement scale, we are directly scaling the range at which the interaction transitions from perturbative to non-perturbative.}
       \label{fig:scaling}
     \end{minipage}\\[1em]
      \end{figure}

       To compute the mass spectra of a dark QCD, we apply the hCQM and scale relevant parameters with the confinement scale. In the simplest potential model with an inter-quark interaction as in Eq. \ref{eq:cornell},
       the size of the bound state can be compared to the radius at which the potential transitions from Coulomb-like to linear. This follows directly from our treatment of the confinement scale as a free parameter, in that we are adjusting the scale
       at which the hidden QCD theory transitions from perturbative to non-perturbative. This transition radius then decreases inversely with $\xi$. This can be seen directly in the case of Eq. \ref{eq:cornell}
       where $k$ has units of $(\text{E})^2$ and so becomes $\xi^2 k$ in a scaled potential. The crossing point is then $r_c=\sqrt{\frac{\tau}{k}}\frac{1}{\xi} $. This relationship can be seen in Fig. \ref{fig:scaling}.
      The shape of the potential directly affects the masses of bound states and in particular has important consequences for resonance states and the size of the hyperfine corrections.
      We now turn to the computation of the dark hadron spectra of different cases of a hidden QCD. The masses also depend on the reference 
      potential that we can scale from and which is taken from past work on potential models of QCD in order to replicate the masses of the hadrons of the standard model.

\subsection{\bf Baryon Spectra}\label{sec:BS}
       We distinguish the cases by the number of light dark quarks in the theory and examine how the spectra change with the confinement scale.
       Including electromagnetic effects we can consider that as $\Lambda_{DM}$ increases and $r_c$ decreases the size of the EM mass contribution to neutral and charged states will be more significant. 
       In particular, if we consider the simple expression for the scaling of the EM self energy of a neutron as \cite{Horowitz:2000kn}
       \begin{equation}
       \label{eq:emcorrect}
        \Delta_{EM} \sim -\frac{\alpha}{\braket{x^2}^{1/2}},
       \end{equation}
       then this term will scale upwards with $\xi$ as the distance becomes smaller. The proton by comparison has a positive mass contribution that also scales with $\xi$ and the difference between these will push 
       charged states above the masses of neutral states for the case of light quarks of equal mass. Such contributions are not however to be subtracted from the calculated neutral ground states in 
       the hypercentral analysis as they are in theory already factored in by the Coulombic term scaling of the potential
       and the fact that the potential was fitted to the experimental neutral masses from the PDG for $\xi=1$. For this reason, our work is most applicable to theories of a dark QCD with an EM U(1) coupling strength the same as that of 
       ordinary electromagnetism. In models of broken mirror symmetries the value of the dark sector's EM coupling constant is constrained to be very close to that of the SM value due to the opposite direction of the running of $U(1)_Y$ and $SU(2)_W$ and so the model
       in this work is directly applicable to the QCD spectra of these models. In a theory without an EM gauge group, the EM mass contribution to the effective potential must be separated in order to remove its effect from the mass ordering.
       
       For larger confinement scales the effect of EM U(1) force in the theory will create mass differences pushing any charged states well above any light neutral ground states if the set of light quarks allows for them. 
       The only counter to this is if the bare dark quark mass differences compensate for the EM mass difference as in the case of the proton-neutron mass splitting of ordinary QCD. 
       This can be contrasted with the effect of $\xi$ on the chromomagnetic spin-spin interactions that we employ and which scale inversely with the dark confinement scale
       and thus increase the degeneracy between the doublet and quartet in two flavour dark isospin, and between the octet and decuplet in three flavour dark isospin. 
       This term crucially depends on the spatial wavefunction and the contact term for overlapping quark coordinates $\braket{\delta^{(3)}} = |\psi(0)|^2$.
       The spin-spin interaction in the form of the chromomagnetic contact term is \cite{DeRujula:1975qlm}
       \begin{equation}
       \label{eq:spinspin}
        V_{ss}= \sum_{i<j} \frac{4}{9 \pi}\frac{\alpha_s}{m_i m_j}\, \delta^{(3)}(r_{ij}) \, \sigma_i . \sigma_j\, .
       \end{equation}
       Note the inverse scaling with constituent masses $m_i m_j$ which will increase degeneracy between the mixed symmetry and totally symmetric baryon multiplets. 
       This term is analogous to the magnetic spin-spin contact interaction that gives rise to the hyperfine splittings in atomic theory however in this case is motivated by the color-magnetic moments. 
       This term is important in understanding the $\Lambda^0 - \Sigma^0$ and $\Delta^0 - N$ mass differences in QCD where the flavour composition is identical and the spin-flavour wavefunction is different, a unique feature for 
       baryon wavefunction ground states when the number of flavours exceeds two.
       They are then similarly important for the present work as they contribute to the mass splitting between the different ground state wavefunctions allowed in the constituent quark model.
       One could also consider the spin-flavour and spin-orbit interactions and depending on the choice of potential these may contribute more or less significantly. We discuss these possibilities further in the appendix.
       In this work we consider models where the spin-spin interaction is the dominant source of these mass differences.
       In the hypercentral assumption with only one hyperradius we lose the ability to directly calculate the full value of the contact term $\braket{\delta^{(3)}}$ for the full coordinate system of three quark wavefunction. 
       The Gaussian-smeared contact term 
       with a functional form 
       \begin{equation}
       \delta^{(3)}(x)= \kappa\,  e^{x^2/r_0^2},
        \end{equation}
       that is treated perturbatively, is an approximation which has been applied successfully 
       to fitting the light baryons in \cite{Varga:1998wp, Giannini:2012vy} among others and we similarly use it in this work for the extrapolation to dark QCD states.
      \\
      \indent  In fitting the form of the potential we compare parameter fits done in similar models for standard 
       hadronic spectra. We consider primarily potentials generalizing that of Eq. \ref{eq:cornell},
       \begin{equation}
        \label{eq:potentialvariation}
         V(x) = -\frac{\tau}{x} + k x^{\rho},
      \end{equation}
       as well as a perturbative hyperfine interaction given by Eq. \ref{eq:spinspin}. The eigenstates are then given by $ E_{N \gamma}$.  
       This follows the work on visible QCD in \cite{Semay:1997ys, SilvestreBrac:1996bg, Varga:1998wp} as well as \cite{Bhaduri:1981pn}. 
       The masses of the baryons are then given by $M_B = E_0 + E_{N L}$ where $E_0 =3 m_q$, i.e. it is the quantity that scales directly with $\xi$ along with the dressed light quark masses.
       With these different potentials, which all fit the experimental spectra to varying degrees, we can examine the variation of dark baryon masses with the choice of potential.  
       Following the potentials given in \cite{Semay:1997ys, SilvestreBrac:1996bg, Varga:1998wp, Bhaduri:1981pn} we list in Table 1 three sets of parameters that with Eq. \ref{eq:potentialvariation} provide a good fit to the hadronic spectra. 
       While our starting point was these potentials, our exact choice of parameters prioritizes the fit to the ground states over the resonances as these are the most relevant to this work. We are then assuming that such a potential, when scaled, provides the more accurate 
       prediction of dark QCD ground states.

                  \begin{table}[h!]
                  \centering
                  \caption{Parameter sets of the three choices of potentials in the fitting to ground states of QCD. The three parameter sets (P1, P2, P3) are taken from the works in \cite{Semay:1997ys, SilvestreBrac:1996bg, Varga:1998wp} respectively. 
                  The choice of units reflects the units used in the original works.} 
                  \label{tab:p1}
                  \begin{tabular}{ |l|l|l|l|l|l|l|l|l|l| }
                  \bottomrule Model &  $\tau$               & $k$            & $\rho$  & $\kappa$        & $m_q$       & $r_0$  \;\;         & $E_0$   \\ \hline
                  P1 &   $102.67 \, MeV.fm$     & $940.95 \, MeV/fm$     & $1$  \;\;   & $616.02 \, MeV.fm$ & $337 \, MeV$   & $0.45\, fm$ \;\;     & $913.5 \, MeV$                                                                              \\ 
                  P2  &    $0.5069$           &  $0.1653 \, GeV^2$     & $1$  \;\;   & $1.8609$        & $0.315 \, GeV$ & $2.3\, GeV^{-1} \, $  \;\;     & $0.8321 \, GeV$                                                                                           \\ 
                  P3  &    $0.4242$           &  $0.3898 \, GeV^{5/3}$  & $2/3$\;\;   & $1.8025$        & $0.277 \,GeV$     & $2.67\, GeV^{-1}$  \;\; & $1.1313 \, GeV$ \\ \hline
                  \end{tabular}
                  \end{table}

  \begin{figure}[]
  \begin{minipage}{\textwidth}
  \centering
  \includegraphics[width=.8\textwidth]{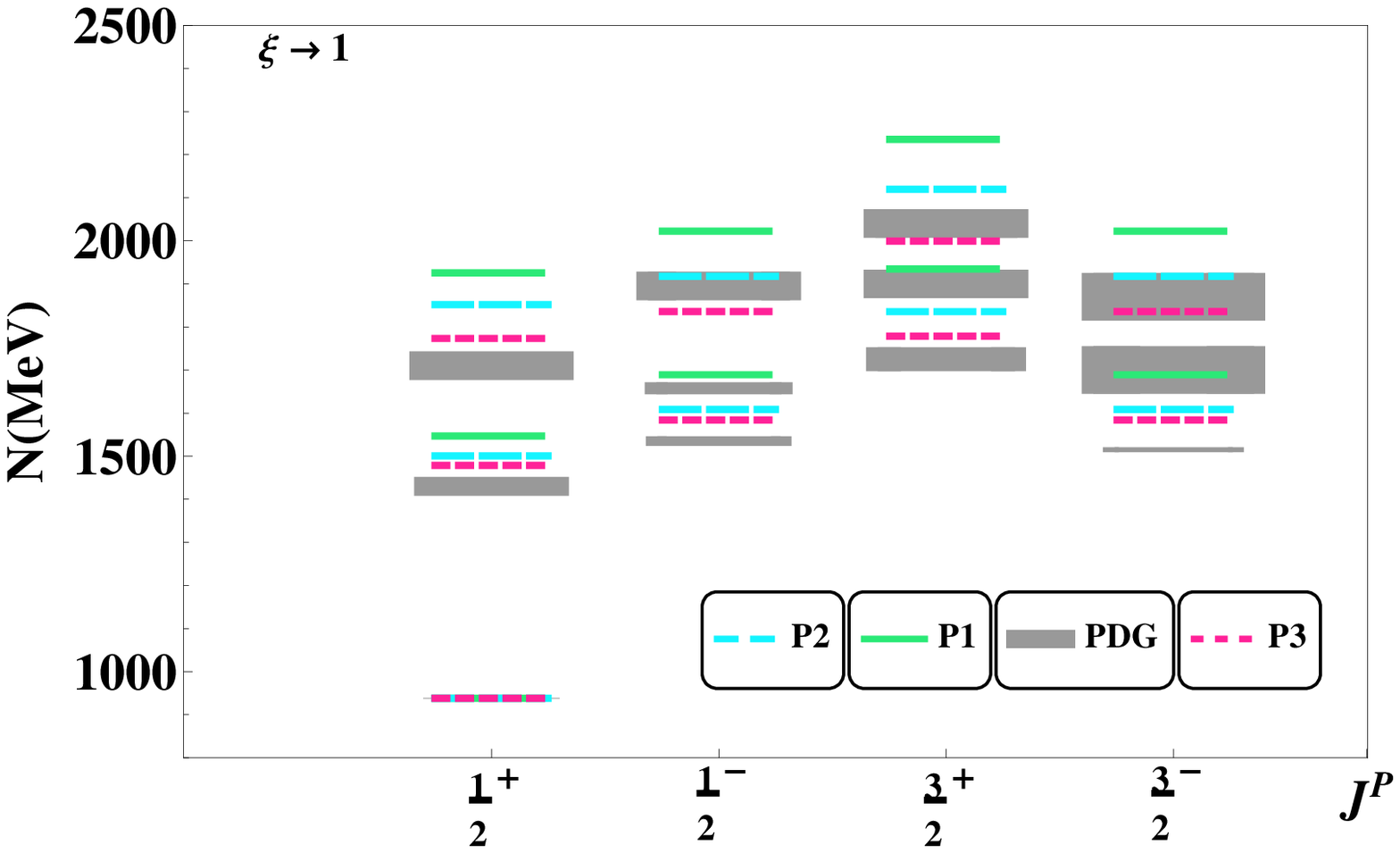}\ %
  \includegraphics[width=.8\textwidth]{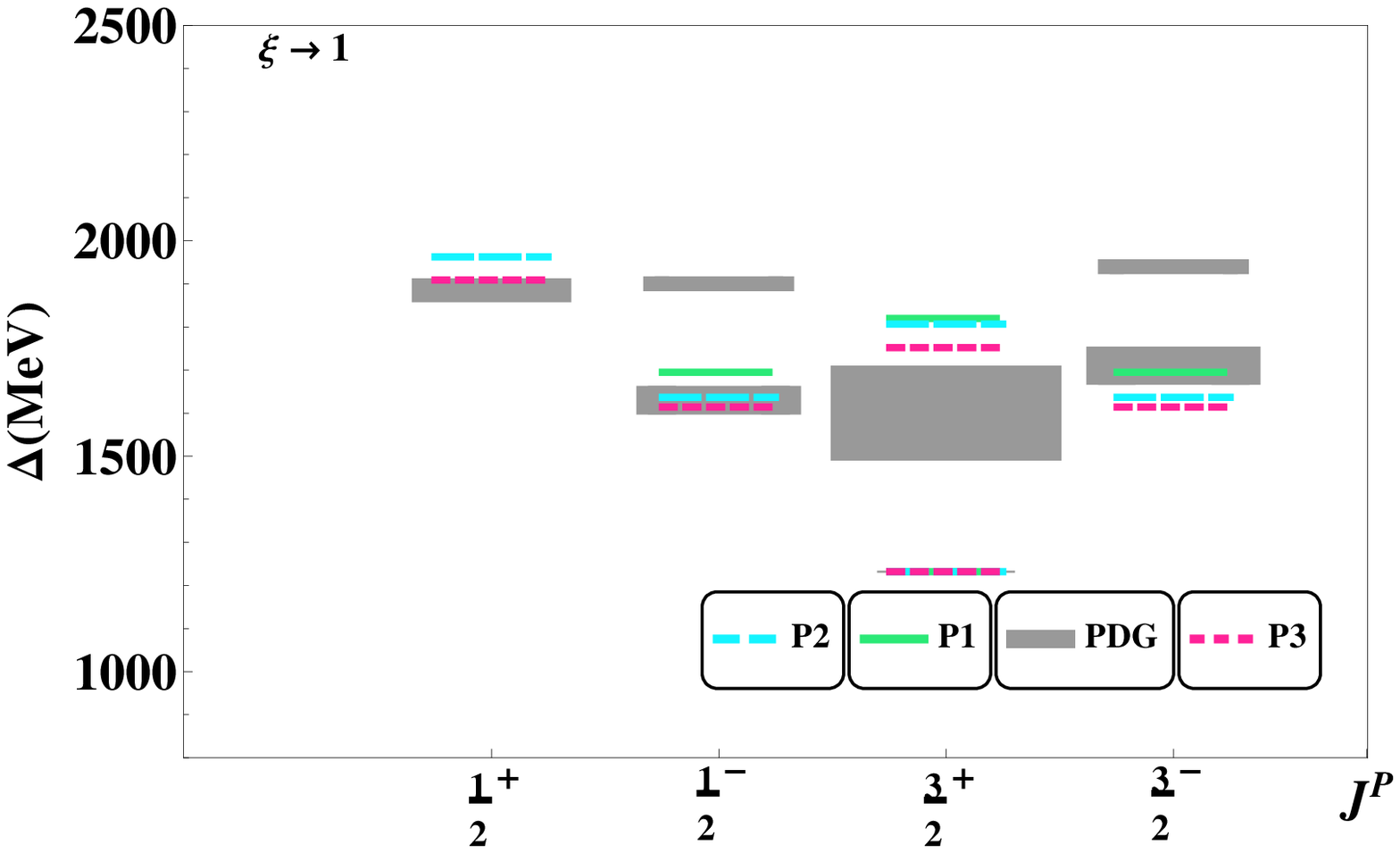}\\   %
  \caption{The resonances of the N and $\Delta^0$ states compared to PDG.
  The spin 1/2 ground state of the neutron and spin 3/2 ground state of the $\Delta^0$ are fitted to match the experimental ground states exactly in each case. These values serve as the reference for dark QCD calculations.   }
  \label{fig:QCDSM}
  \end{minipage}\\[1em]
  \end{figure}

\begin{figure}[t]
    \begin{minipage}{\textwidth}
    \centering
    \includegraphics[width=.8\textwidth]{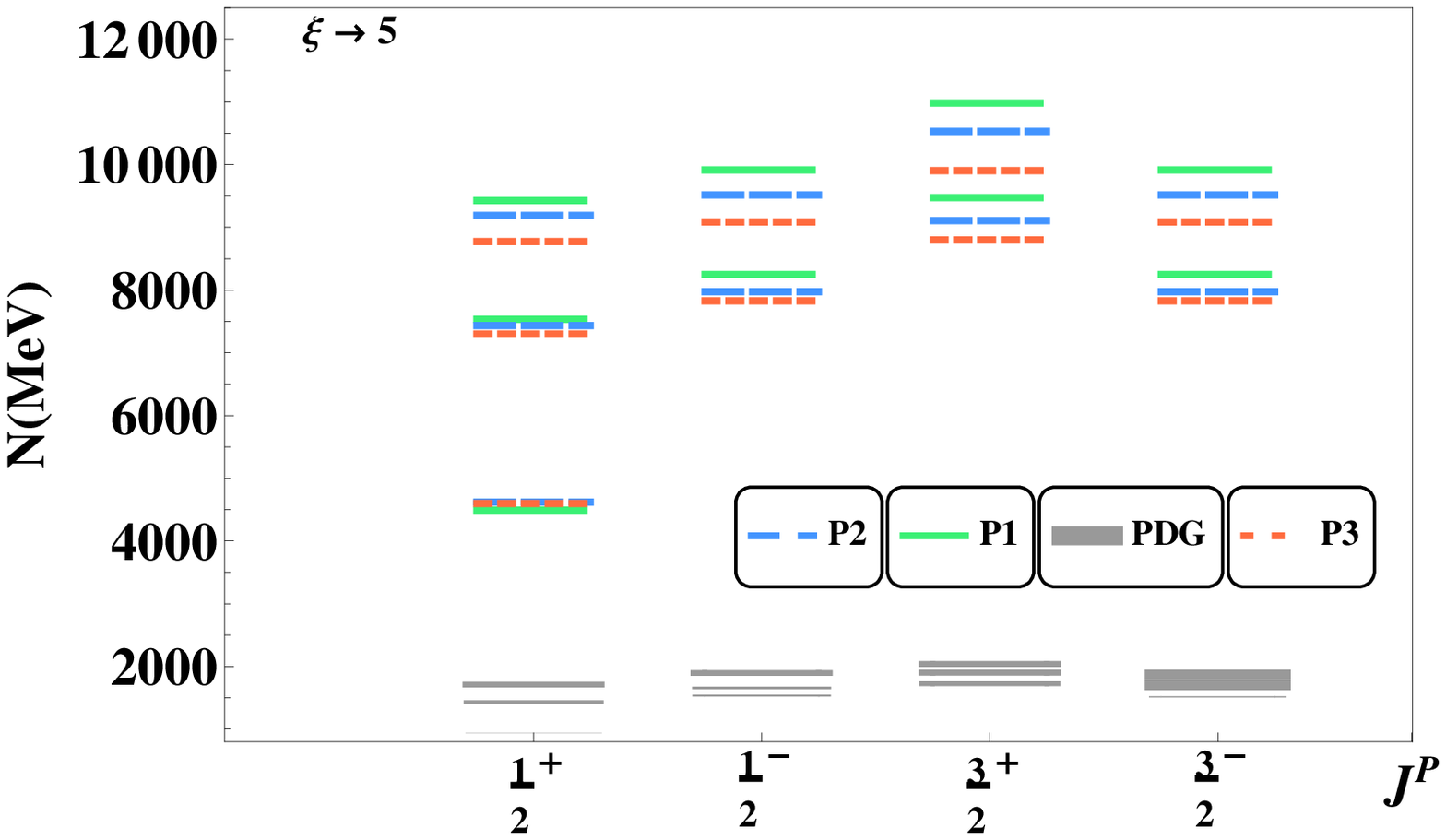}\ 
    \includegraphics[width=.8\textwidth]{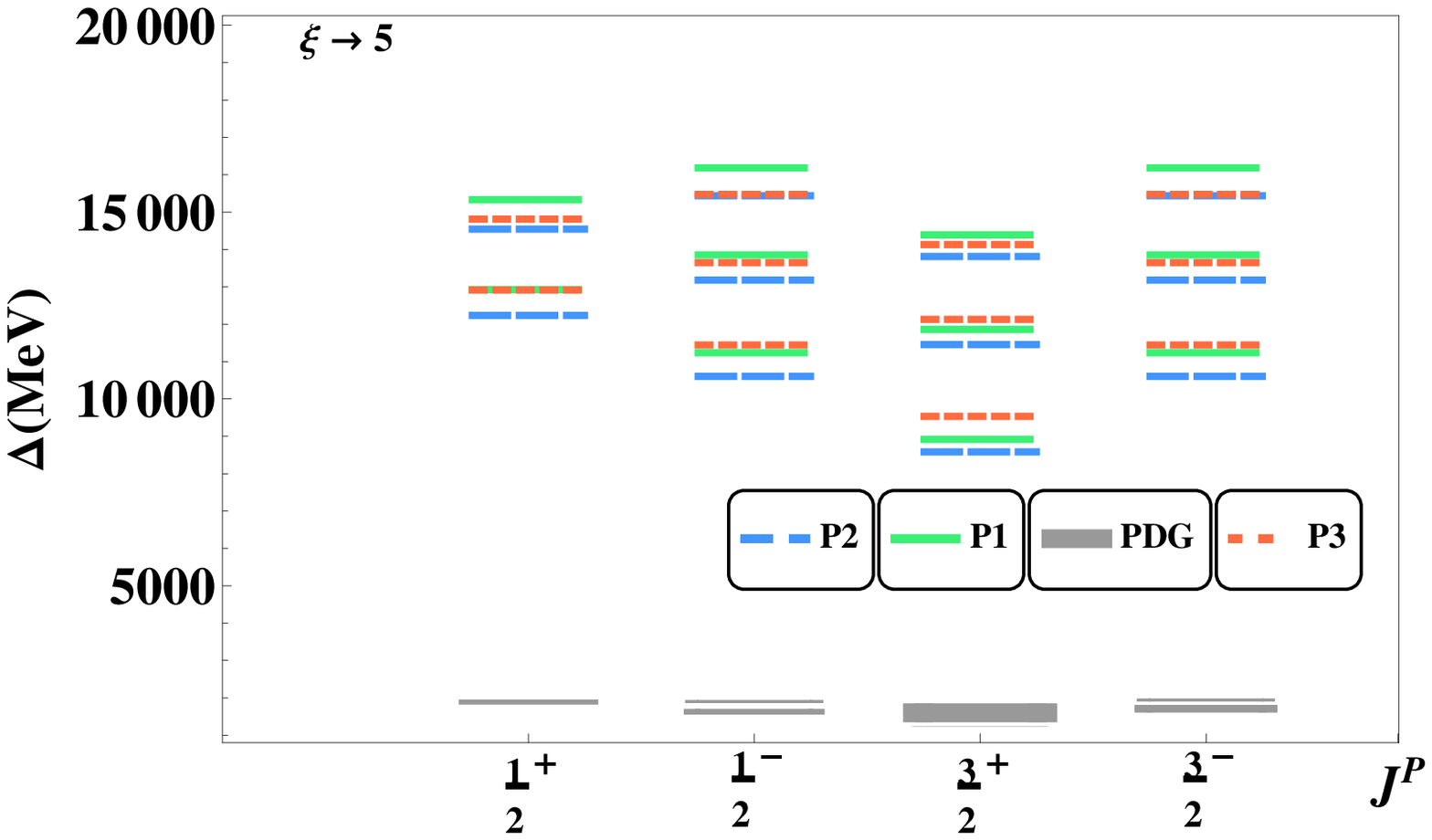}\\      
    \caption{Spin 1/2 and spin 3/2 for a confinement scale ratio $\xi=5$ and three light quarks. Dimensional parameters scale with $\xi$ displayed in the upper left. 
    The PDG values for the experimental spectra are shown for reference to the scale of the $\xi=1$ case. Each of the results (P1,P2,P3) corresponds to a choice of parameters for the $\xi=1$ potential given in Table 1. }
    \label{fig:dQCDndelta}
    \end{minipage}\\[1em]
    \end{figure}

       We can then calculate the same states for a dark QCD model as a function of the number of light quarks and a value of $\xi$. The parameters ($\tau,k,\rho,\kappa, m_q, r_0, E_0$) then all 
       scale appropriately according to the mass dimensions of the chosen potential as discussed in Fig. \ref{fig:scaling}. Figure \ref{fig:QCDSM} shows the fit to ground states of the $N$ and $\Delta$ baryons in QCD while 
       Fig. \ref{fig:dQCDndelta} shows a scaled version for a value of $\xi=5$, chosen for the sake of example.
       \\
       \indent In the one flavour quark case, the baryon spectrum consists of a lightest stable $\Delta$ baryon with ground state spin 3/2. 
       In the case of dark electromagnetic $U(1)_{Q}$ symmetry consistent with the SM it would have EM charge +2 in the case of a single up type quark. 
       It could also be a single down type and so be singly charged with opposite sign. For the mass of this state we can compare with standard QCD in that we 
       calculate the mass from the constituent quarks and the potential energy from the scaled potential including the spin interactions that lifts the ground state
       according to the chromomagnetic hyperfine interaction in Eq. \ref{eq:spinspin}. 
       The mesonic states will likewise contain only one state however this lightest meson will be unique in that it has the feature of an anomalous mass from the breaking of the anomalous axial U(1).
       The size of this anomalous mass in these models of dark QCD with one light flavour is beyond the scope of this work.
       \\\\%
       In the two flavor quark case, with an isospin symmetry among the light states, the baryon sector will have a spin-flavour SU(4) symmetry. The spectrum then consists of a 
       $\Delta$ quartet of spin 3/2 states as well as a spin 1/2 pair $(N,P)$. However we can also consider the cases where this doublet and quartet have charges that 
       follow the possibility of the two light dark quarks being both up type or both down type.  
       The splitting between the doublet and quartet in any scenario is modeled again with the spin-spin contribution which gives the spin 3/2 states 
       larger mass than similar spin 1/2 states while EM effects will make the neutral states lighter in general. For sufficiently light quarks with near equal masses we can consider the spin 1/2 (N,P) doublet as the 
       lightest states in the first case, and with $U(1)_Q$ corrections selecting the neutral state as the lightest ground state. 
       In the case of two up-like or down-like dark quarks, we have degenerate states with equal EM contributions which we 
       label ($\Sigma^{++}_c,\Xi^{++}_c$) and ($\Xi^{-}, \Sigma^{-}$) following the naming conventions of 
       standard QCD where in this case the flavour content of the theory is taken to consist of near degenerate (u,c) or (d,s) dark quarks. 
       \\\\%
       In the three flavour case we can compare directly with QCD, however again we can distinguish the cases according to other 
       quantum numbers. With three flavours we recover the familiar octet and decuplet however with near degenerate quark masses
       the spectrum will be near degenerate in flavour unlike the strange quark mass splitting seen in QCD. Again we can consider EM mass differences where flavour content allows for neutral and charged states. 
       In the case of three or more flavours we gain two ways of forming a spin 1/2 wavefunction for ground states as in the case of the $\Sigma, \Lambda$. 
       The differences in terms of Eq. \ref{eq:spinspin} are the values of the $\sigma.\sigma$ terms. It remains true however that in the degenerate u,d,s case that $N ,\Sigma, \Lambda$ have the 
       same mass and share the place of the lightest state.

\begin{figure}[h!]
    \begin{minipage}{\textwidth}
    \centering
    \includegraphics[width=.75\textwidth]{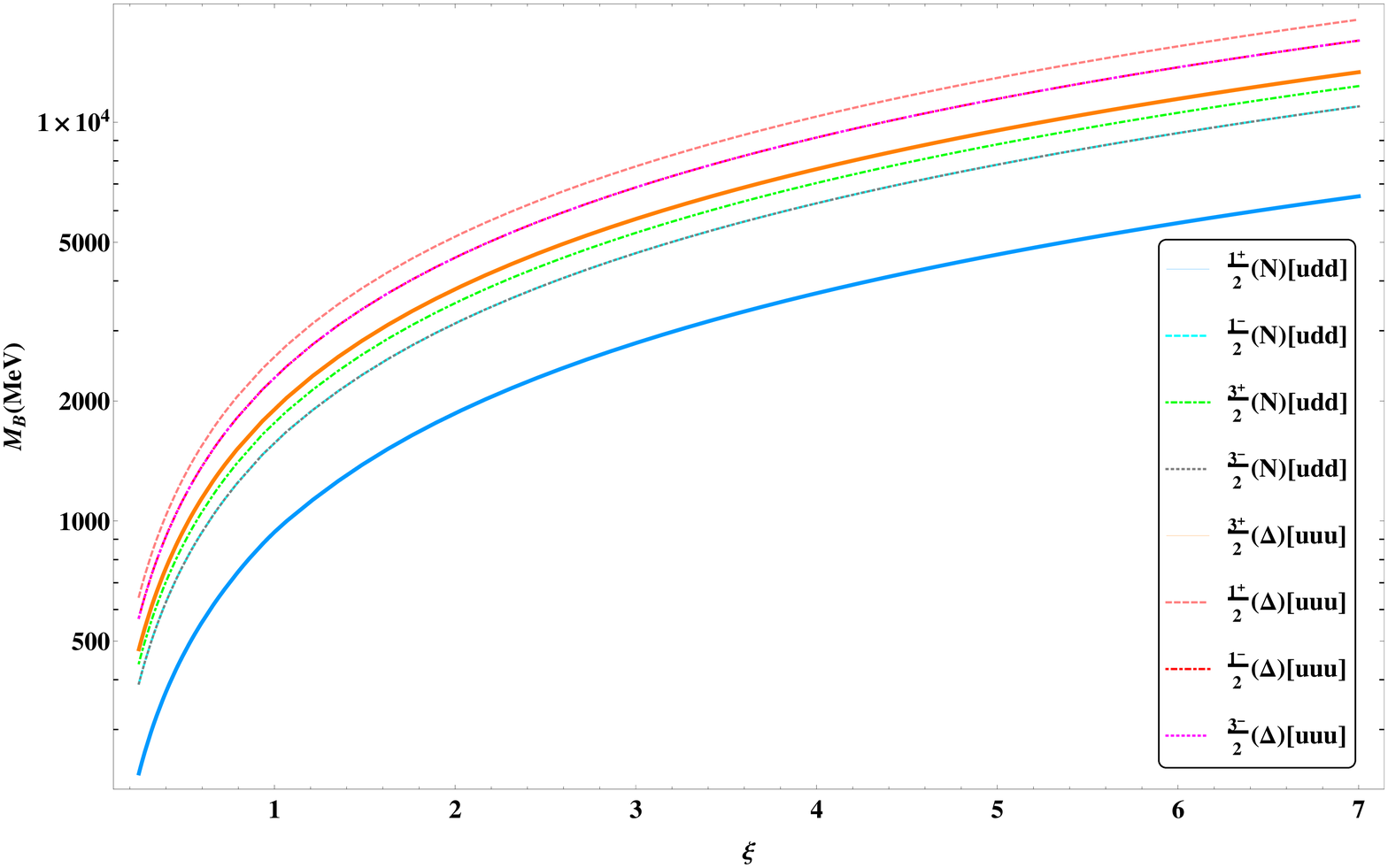}\ 
    \caption{Spin 1/2 and spin 3/2 states as a function of the parameter $\xi$, the ratio of confinement scales. The direct scaling with confinement scale becomes more significant at large $\xi$ as one expects from dimensional analysis.
    This uses an average of the three parameter regimes P1,P2 and P3. Using this plot we can see at each value of $\xi$ the lowest lying baryon in a dark QCD spectrum for models of: one flavour (uuu, orange), or with two or more flavours (udd, blue).}
    \label{fig:runrun}
    \end{minipage}\\[1em]
    \end{figure}

       In Figure \ref{fig:runrun} we show as a function of $\xi$ the lightest spin parity states for each of the lightest baryons in the cases of only one light quark and two light quarks. The case of two light quarks however provides the mass scale
       of the lightest state for any number of light quarks greater or equal to two.

  \begin{figure}[h!]
    \begin{minipage}{\textwidth}
    \centering
 \includegraphics[width=.75\textwidth]{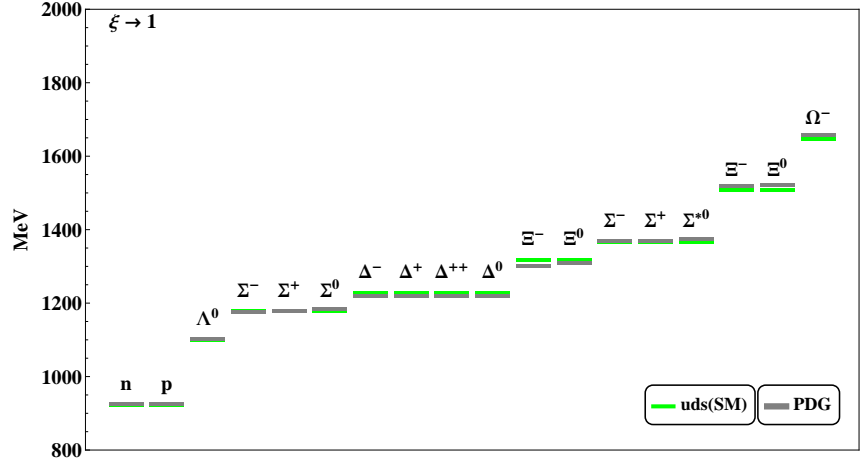}\\ 
    \caption{Baryon states in our model for visible QCD compared to PDG values \cite{Agashe:2014kda}. Parameters for the G{\"u}rsey-Radicati formula are fitted from the baryon mass results of the hypercentral Schrodinger equation.}
    \label{fig:GRPDG}
    \end{minipage}\\[1em]
    \end{figure}

         \begin{figure}[h!]
    \begin{minipage}{\textwidth}
    \centering
    \includegraphics[width=.49\textwidth]{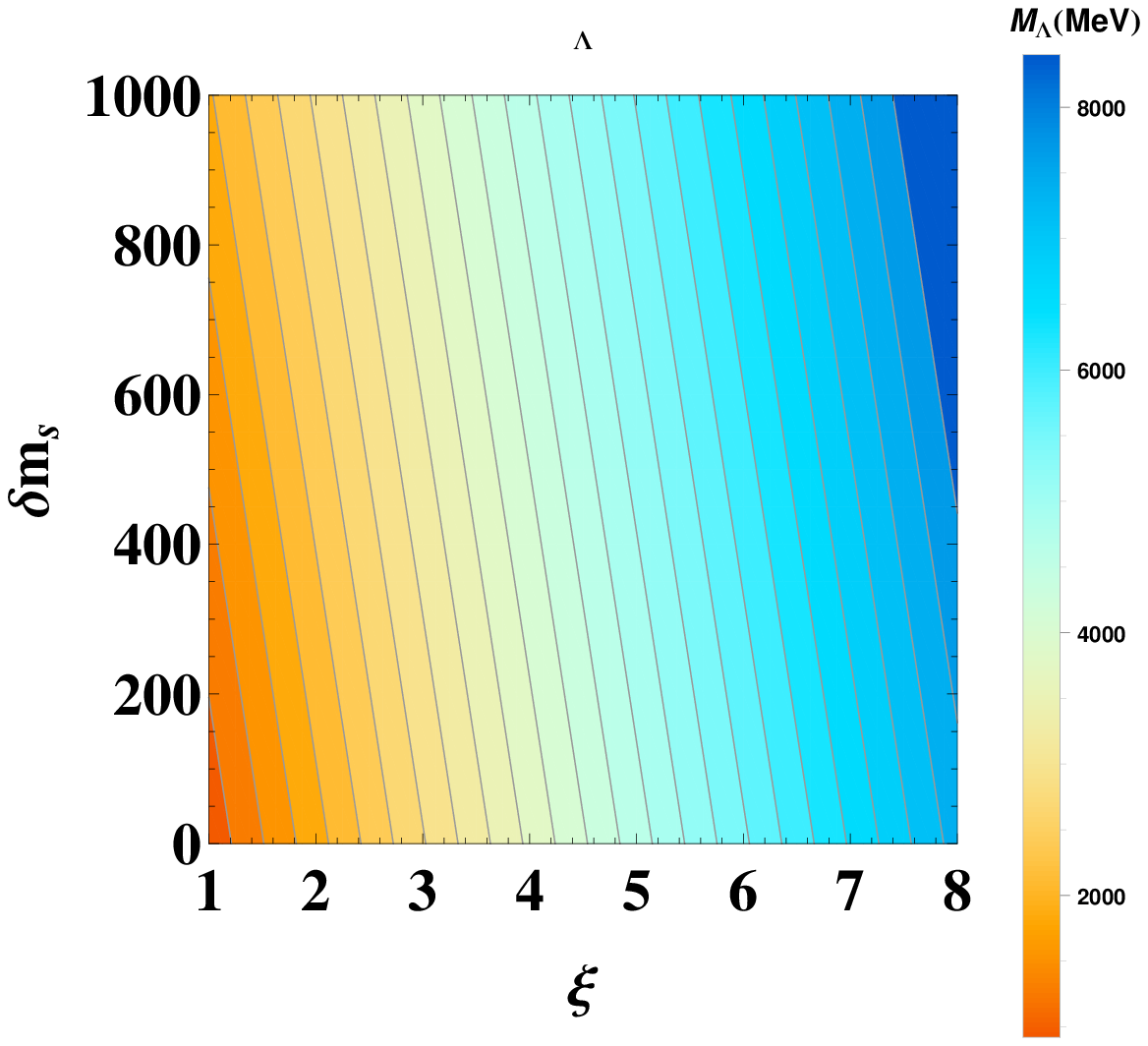}\ 
    \includegraphics[width=.49\textwidth]{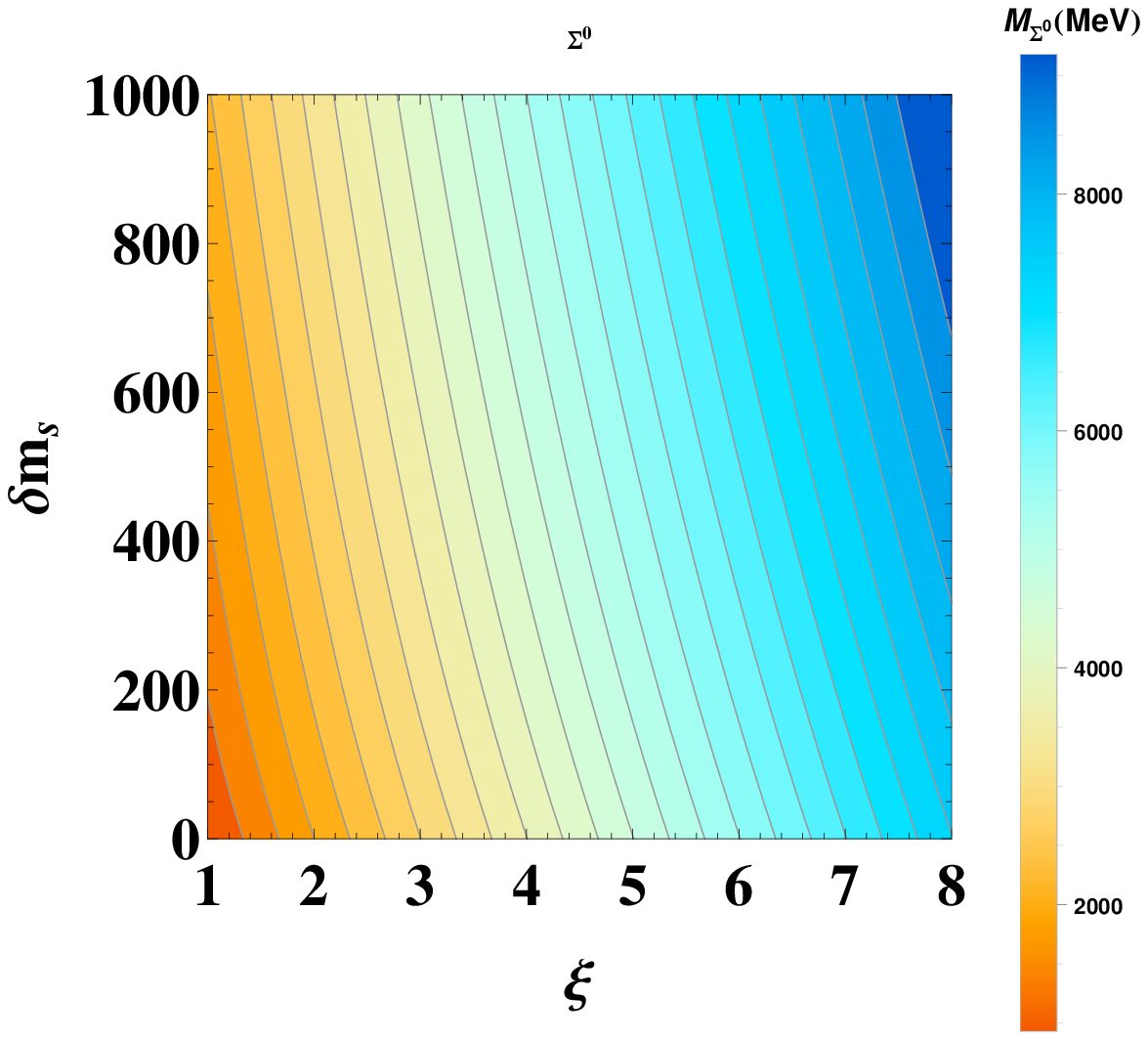}\\      
        \includegraphics[width=.49\textwidth]{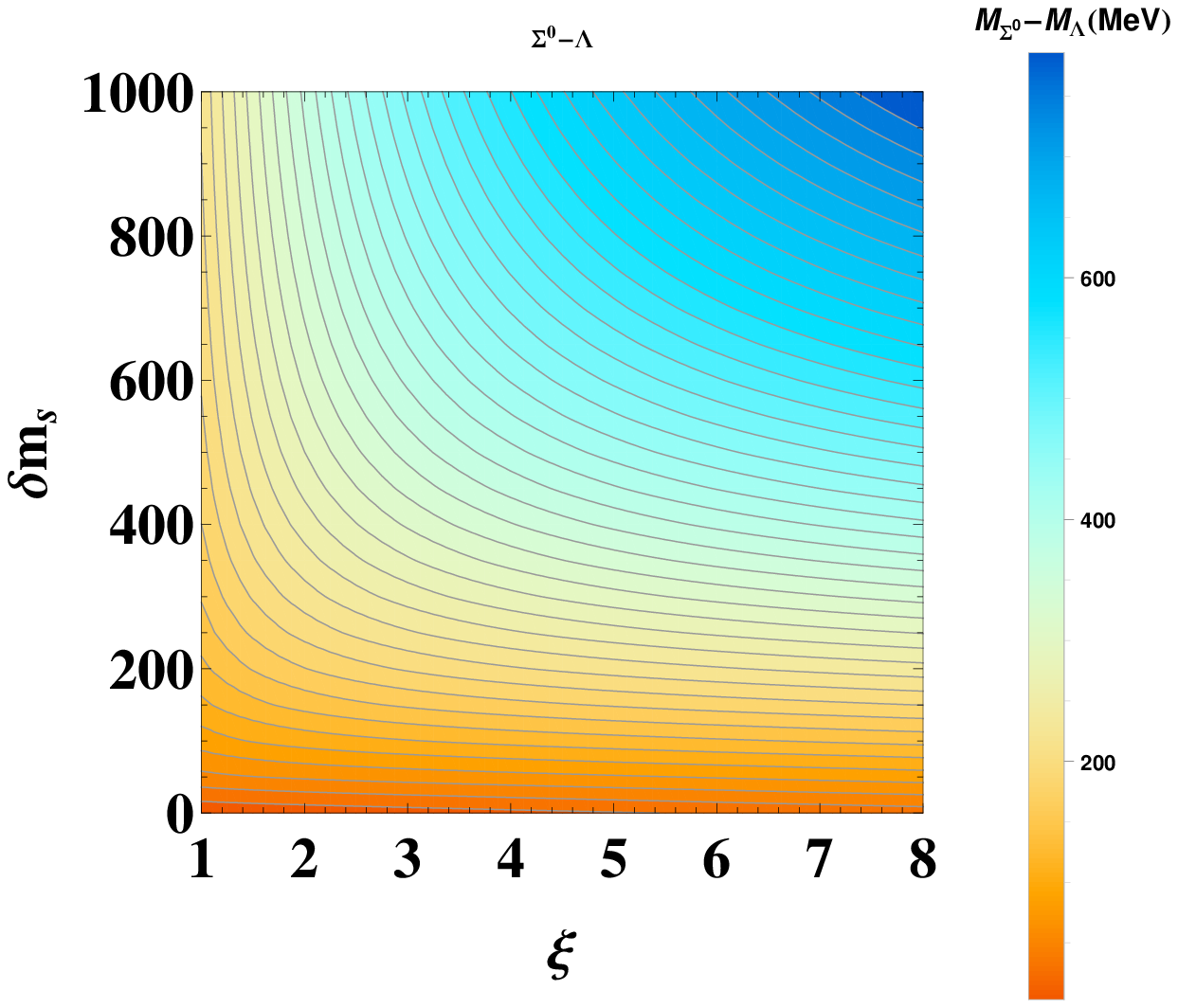}\     
      \includegraphics[width=.49\textwidth]{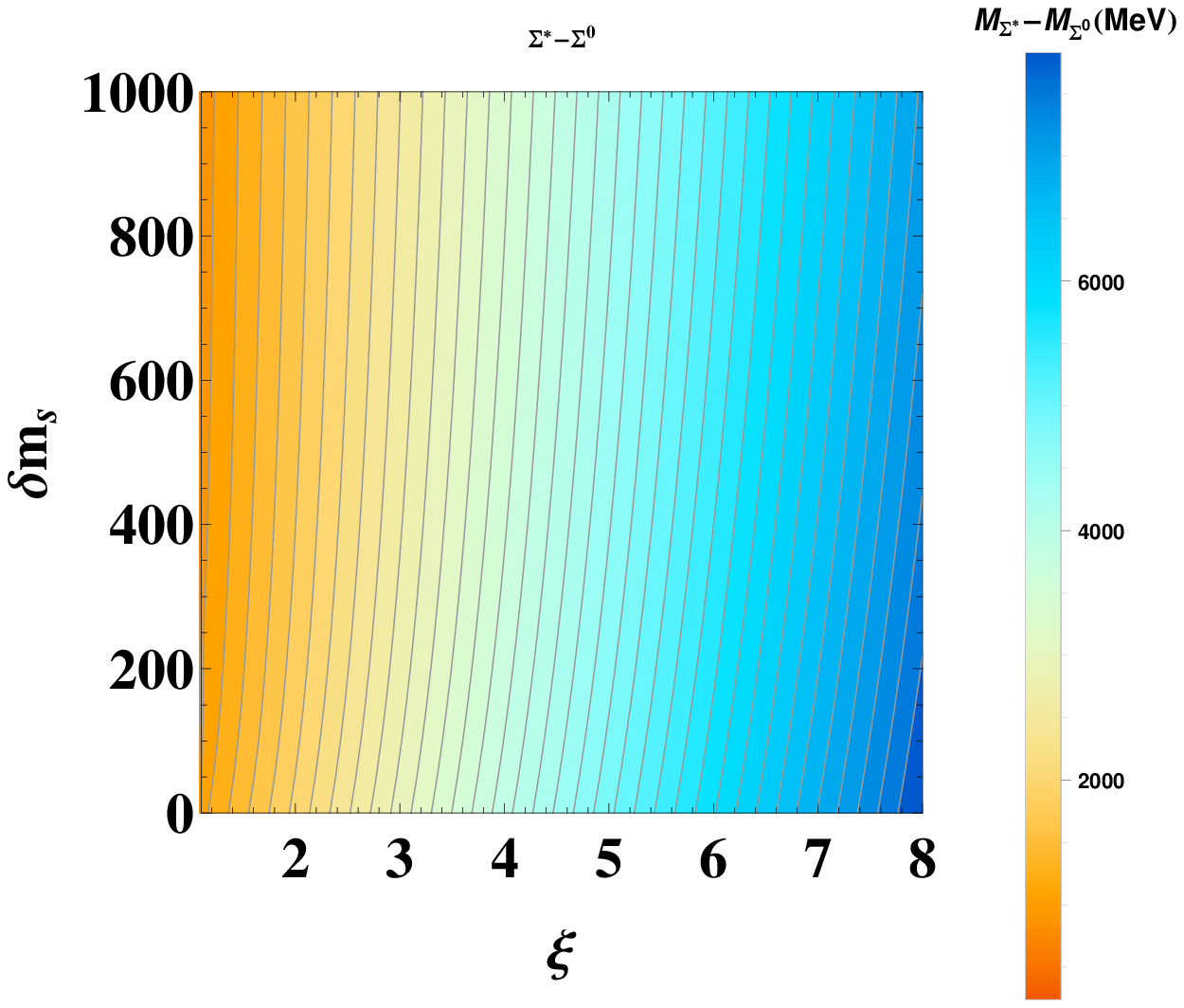}\     
    \caption{The top two plots shows the masses of the dark sector $\Lambda^0$ and $\Sigma$ and states for a range of values of the semi light mass splitting $\delta m_{s}$ and $\xi$. The lower plots display the mass differences for the $\Sigma$ and $\Sigma^*$ baryons as well as the mass splitting 
    of the two different 3-quark spin-flavour wavefunctions ($\Lambda$, $\Sigma$) for a range of $\delta m_{s}$ and $\xi$.}
    \label{fig:conn}
    \end{minipage}\\[1em]
    \end{figure}

\begin{figure}[h!]
       \begin{minipage}{\textwidth}
       \centering
       \includegraphics[width=.98\textwidth]{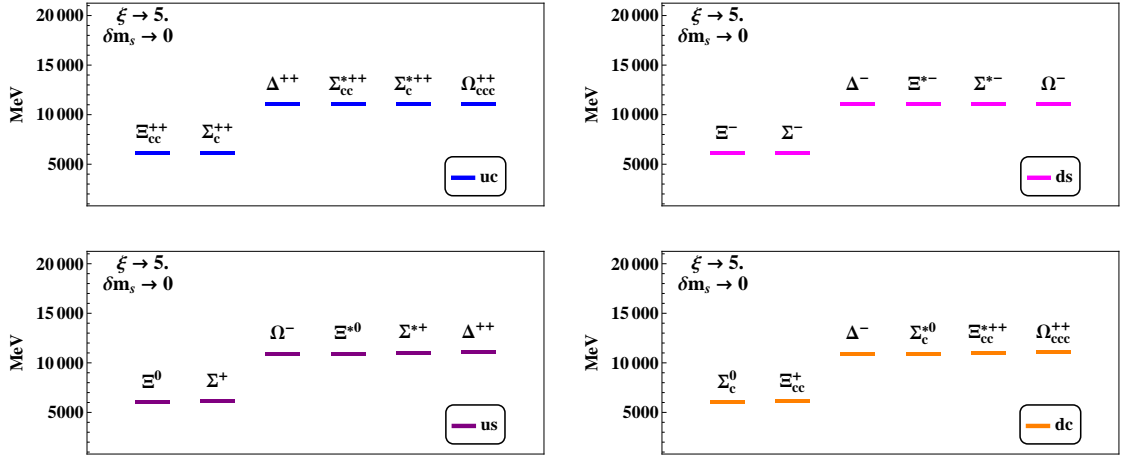}\quad 
       \caption{G{\"u}rsey Radicati states for a hidden QCD with two flavours, $\xi=5$ and degenerate light quark masses in increasing mass values in four flavour combination cases. }
       \label{fig:GURSEYSM3}
       \end{minipage}\\[1em]
       \end{figure}

\begin{figure}[h!]
    \begin{minipage}{\textwidth}
    \centering
    \includegraphics[width=.75\textwidth]{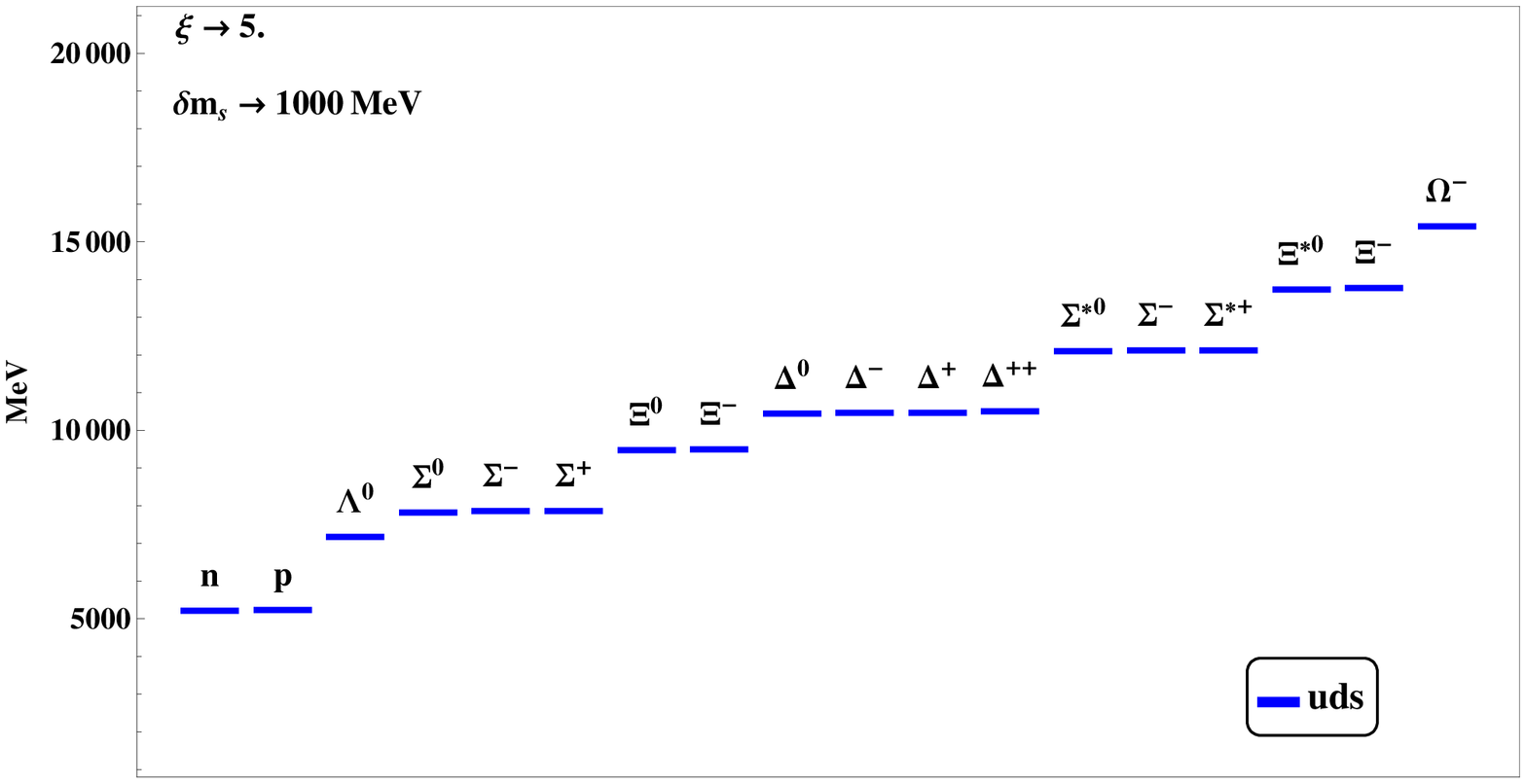}\\    
    \includegraphics[width=.75\textwidth]{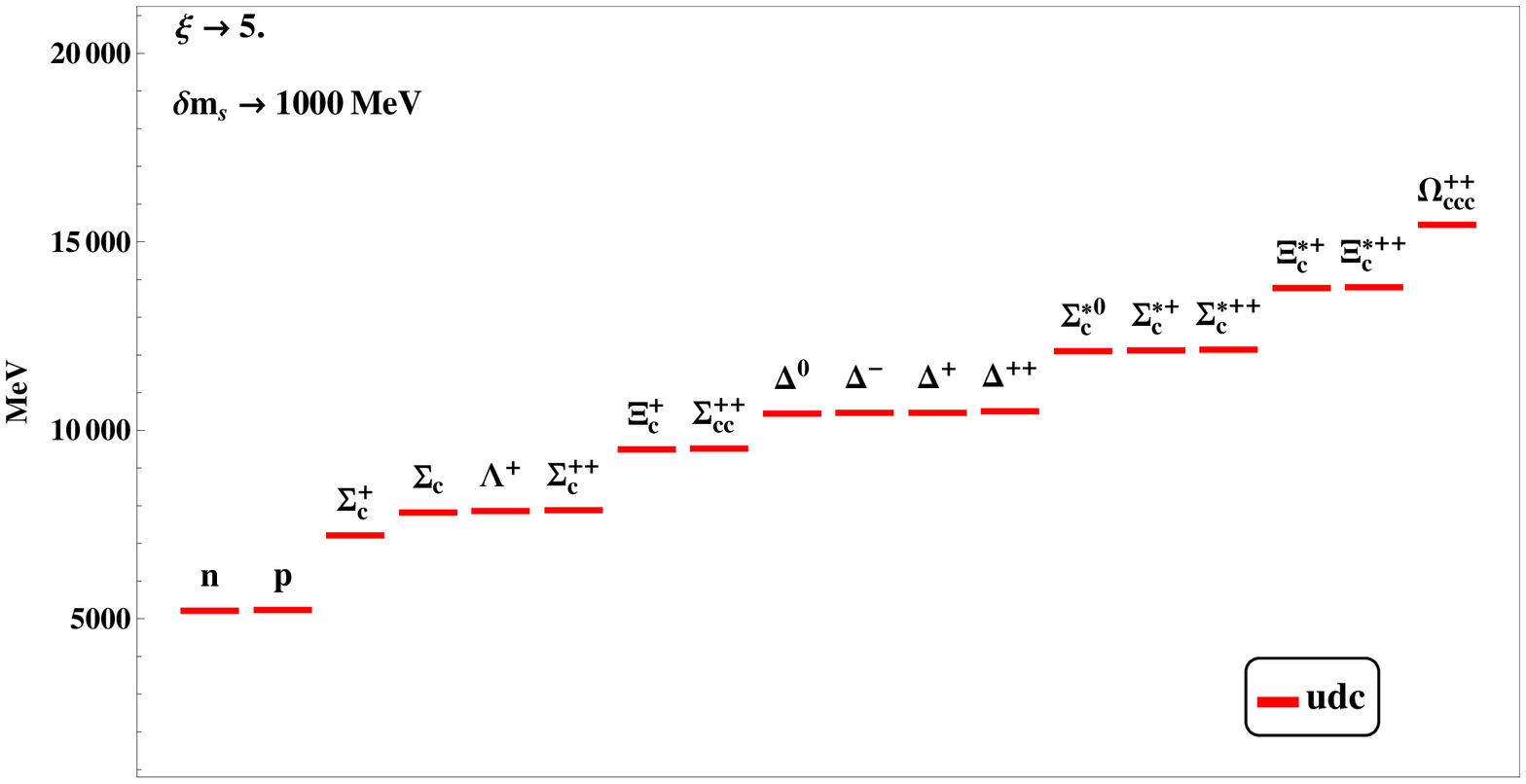}\\ 
    \includegraphics[width=.75\textwidth]{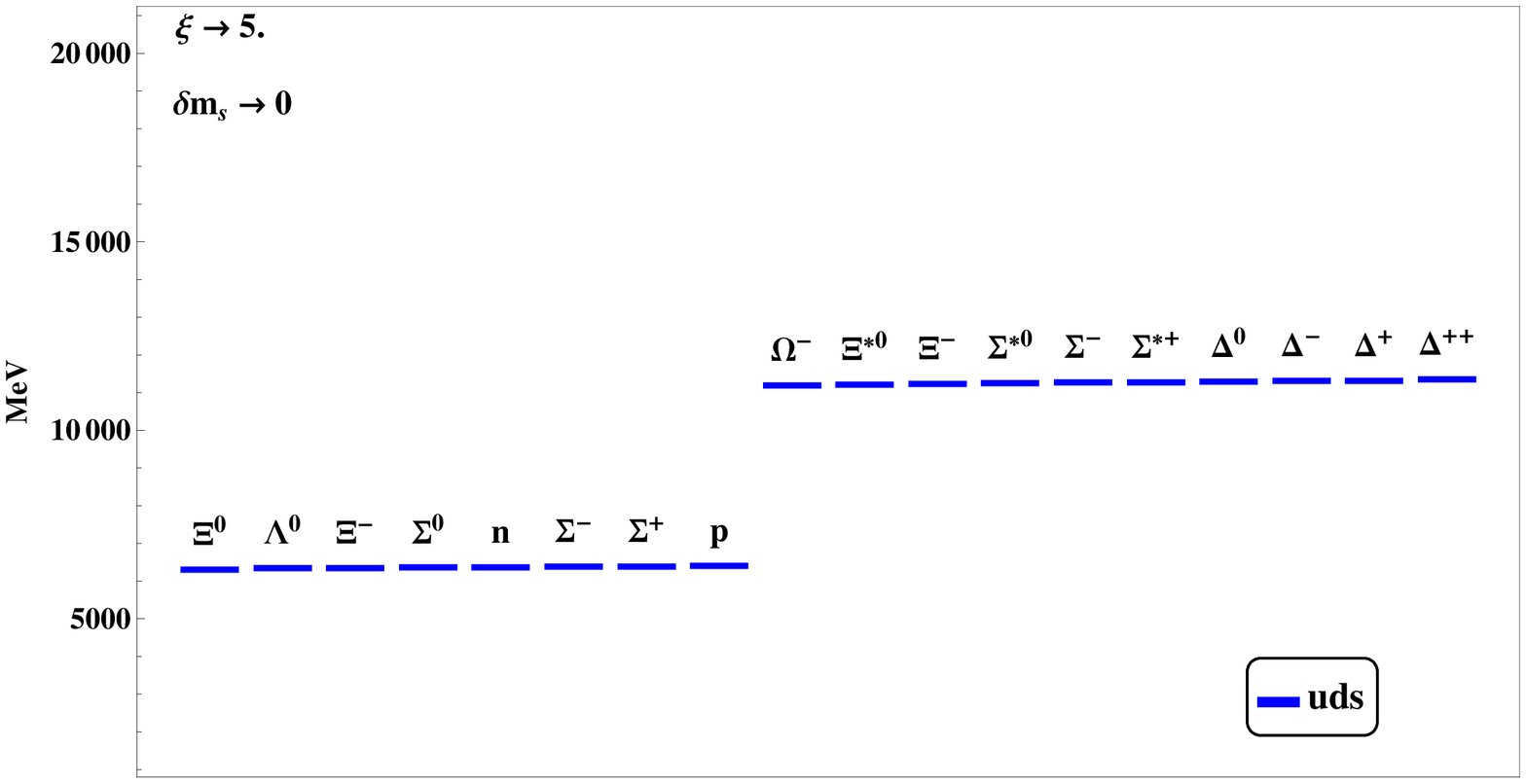}\\      
    \caption{Spin 1/2 and spin 3/2 baryon ground states for a factor of $\xi=5$ in the case of three light quarks where one quark may have additional mass. The more massive state is then labeled as the c or s dark quark depending on its quantum charges.}
    \label{fig:GRX}
    \end{minipage}\\[1em]
    \end{figure}

       By observing the spin and flavour symmetries of a given set of light quarks one can construct an 
       equation with a functional form similar to that of the Gell-Mann-Okubo equation which was used to make sense of the mass splittings of ground state hadrons in the quark model.  
       For a dark variant of such an equation, the unknown parameters cannot be extracted from experiment though we can consider
       how these parameters change from ordinary QCD for a choice of dark QCD. We introduce the additional free parameter $\delta m_s$ as the mass difference between the constituent quark mass of the light states 
       and the constituent mass of a semi-light state, analogous to the position of the strange quark in QCD. In particular we consider the spin, and hypercharge symmetries of the set given by the $SU_S(2)$ and $U_Y(1)$ groups. 
       We also include the generalization of isospin, $SU_I(n_l)$ where $n_l$ is the number of light flavours.
       This leads us to the form of the G{\"u}rsey-Radicati mass formula, which was used to explain the Gell-Mann-Okubo mass relations \cite{Gursey:1992dc}, which is written as 
        \begin{equation}
        M = M_0 +  C\, C_2 \left[SU_{S}(2)\right]+ D\, C_1 \left[U_{Y}(1)\right]
        + E\,\left( C_2 \left[SU_{I}(n_l)\right] -\frac{1}{4}(C_1 \left[U_{Y}(1)\right])^2 \right),
       \end{equation}
       where $C_1 ,C_2$ are the quadratic Casimirs for each group.
       In the case of QCD this becomes
       \begin{equation}
       \label{eq:GR1}
        M = M_0 + C\, S (S+1) + D\,Y + E\,\left[T(T+1) - \frac{1}{4} Y^2\right]
       \end{equation}
       and works quite well in reproducing the masses of the octet and decuplet, of ground states in QCD, as shown in  Fig. \ref{fig:GRPDG}.
       In Eq. \ref{eq:GR1} $M_0$ is a new scale that places the energy of the full baryon spectrum rather than being $E_0$ itself. However from the hypercentral analysis to find the ground state of
       a confining theory with $n_l$ light flavours and a given $E_0$ we can predict the mass of the ground state and working from this result determine the value of $M_0$.
       This idea follows the applications to the experimental QCD spectrum in \cite{Giannini:2005ks} where the SU(6) 
       spin-flavour symmetric Hamiltonian is solved numerically to find the central values for the G{\"u}rsey-Radicati formula. 
       The experimental states typically chosen to fit the parameters are, in terms of ground states,
       \begin{align}
        \label{eq:fitsGR}
         \Sigma^{*}-\Sigma       &= 3C \\ \nonumber
        \Sigma-N   &= \frac{3}{2}E - D \\ \nonumber
        \Lambda-N  &= -D-\frac{1}{2}E. 
        \end{align}
       We can then use Eq. A5 to obtain a minimal number of baryon ground states and
       refit the above parameters, for a choice of $\xi$ and $\delta m_s$, based on our calculated eigenvalues instead of the experimental spectrum. 
       Figure \ref{fig:conn} shows the masses of the $\Lambda^0$ and $\Sigma$ ground states, as well as the mass differences in spin 1/2 and spin 3/2 $\Sigma$ ground states, needed to find the parameters of such
       a formula for a choice of $\xi$ and $\delta m_{s}$ while Figure \ref{fig:runrun} shows the lightest $\frac{1}{2}^+$,$\frac{1}{2}^-$ and $\frac{3}{2}^+$ values needed in order to fit the mass difference parameters
       that use the the neutral N and $\Delta$ states.
       Note that as $\delta m_s$ approaches zero, the dimensional parameters D, E also approach zero, as we expect. This is relevant to the limit of maximum degeneracy.
       It is through this method that the mass differences within the baryon multiplet for a dark QCD model can then be explored by solving for the  G{\"u}rsey-Radicati formula parameters each
       time we generate the resonance spectrum for dark QCD states. 
       Figure \ref{fig:GURSEYSM3} examines the cases with only two light flavours and in Fig. \ref{fig:GRX} we examine how the spectra of lightest spin-flavour states changes depending on 
       the value of $\xi$ and whether there are any non-degenerate light quarks.
       Applying this methodology to a dark QCD inherently comes with the caveats that the exact 
       scaling of these parameters in, for instance, one flavour dark QCD may be more complicated than the scaling we employ. 
       In particular the relationship between bare quark mass and constituent quark mass is non-trivial and has been explored in lattice studies such as \cite{Biernat:2014eya}.

\newpage
\subsection{\bf Meson Spectra}\label{sec:MS}
\begin{figure}[h!]
   \begin{minipage}{\textwidth}
   \centering
   \includegraphics[width=.8\textwidth]{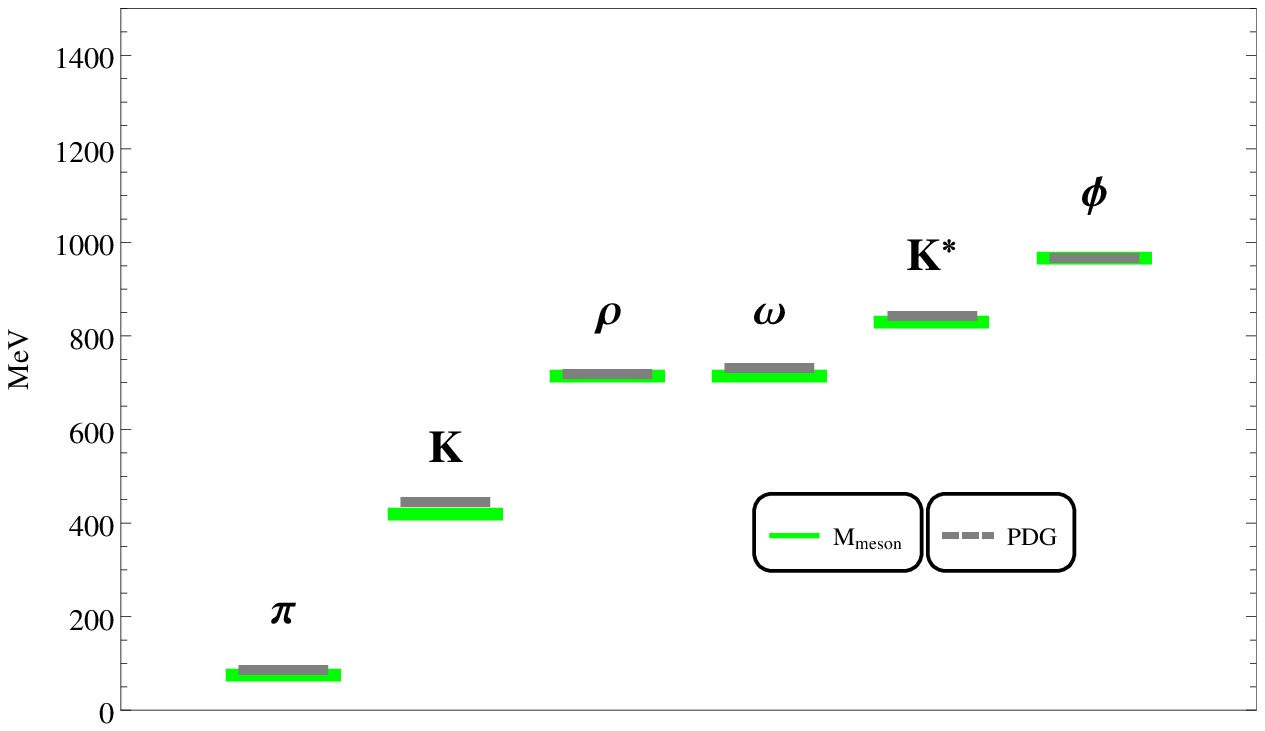}\quad 
   \caption{Meson states from Eq.\ref{eq:messtep} compared to PDG values in QCD. As with the baryon model, these fits will act as the starting point for the scaling of dark meson masses.}
   \label{fig:mesonstep}
   \end{minipage}\\[1em]
   \end{figure}

For mesons we are mostly interested in the scaling with confinement scale as their masses can have significant consequences on the stability of baryons. 
They may additionally impact the cross sections of strong interactions and thus the self-interaction strength of dark matter. In QCD, the application of the constituent quark model meson formula \cite{bookMann},
\begin{equation}
\label{eq:messtep}
 M_{\text{meson}}= m_1+ m_2 + \frac{1}{3}\left(\frac{8 \pi}{3}\right)\frac{4 \pi \alpha_s }{ m_1 m_2} S_1.S_2 |\Psi_{\text{meson}}(0)|^2,
\end{equation}
works surprisingly well, where $S.S$ is 1/4 for vector mesons and -3/4 for pseudoscalars. Taking the up and the down to have constituent mass 310 MeV and the strange quark 
to have 483 MeV reproduces the results in Fig. \ref{fig:mesonstep}. One approach we can then take in exploring a dark analogue is to simply 
consider the inter quark potential from the baryon spectrum and solve for the two body wavefunction to find $ |\Psi_{\text{meson}}(0)|^2$. This follows the work in \cite{Semay:1997ys, SilvestreBrac:1996bg, Varga:1998wp} where 
the potential and parameter space considered was specifically designed to fit both the baryon and meson spectra.
   \

While Eq. \ref{eq:messtep} is not particularly accurate in the chiral limit as $m_q \rightarrow 0$, we observed that the pion scaling with $\xi$ is consistent with the
Gell-Mann, Oakes, Renner  relation,
\begin{equation}
 m^2_{\pi}= \frac{(m_u + m_d) \rho}{f^2_{\pi}},
\end{equation}
if one assumes a pion decay constant that does not vary with $\xi$.
The parameter $\rho$ is the condensate, $\rho=\braket{\bar{q}q}$, that scales directly with $\xi$. 
This suggests that the model we employ is taking the 
degree of explicit chiral symmetry breaking to be of the same magnitude as standard model QCD as we increase $\xi$. 
In other words, while the bare light dark quark masses remain small compared to 
$\Lambda_{DM}$, we are able to analyze the meson spectrum for models where the ratio of light 
dark quark current masses to $\Lambda_{DM}$ is similar to standard QCD. 
This is consistent with models of broken mirror symmetries where the dark EW scale can be a free parameter. This limits the amount of parameter space we can explore in this particular approach to dark hadronic spectra to models
of dark QCD with a similar degree of chiral symmetry breaking.
We additionally know that in the chiral limit the mesons approach zero mass and so Eq.\ref{eq:messtep} is 
applied in the context of increasing $\xi$ with meson spectra fitted dressed masses now scaling with $\xi$ along with the inter-quark potential. We can also factor in a value 
of $\delta m_s$ to observe the splitting with a semi-light dressed state. The meson spectra are much more sensitive to the masses of the bare dark quarks, which are all free parameters, and thus the spectra and the comparison between meson and baryon mass
will depend on the exact model. As noted in \cite{Antipin:2015xia}, choosing a semi-light mass for bare quark 
masses and a hidden QCD without flavour violating dark electroweak forces, one can make states similar to the $\Lambda$ 
baryon stable as Kaons may be too heavy for kinematics to allow decay to lighter baryons.

    \begin{figure}[h!]
  \begin{minipage}{\textwidth}
    \centering
    \includegraphics[width=.47\textwidth]{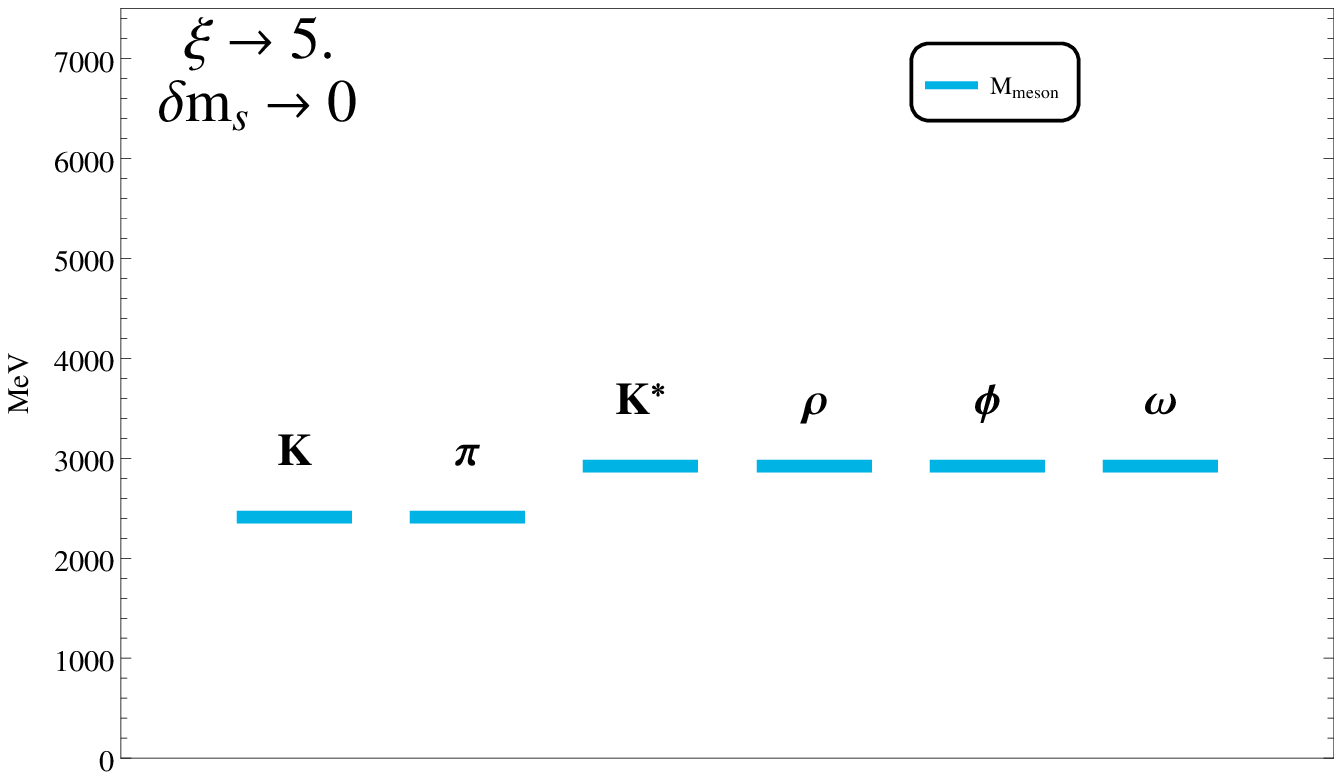}\quad
    \includegraphics[width=.47\textwidth]{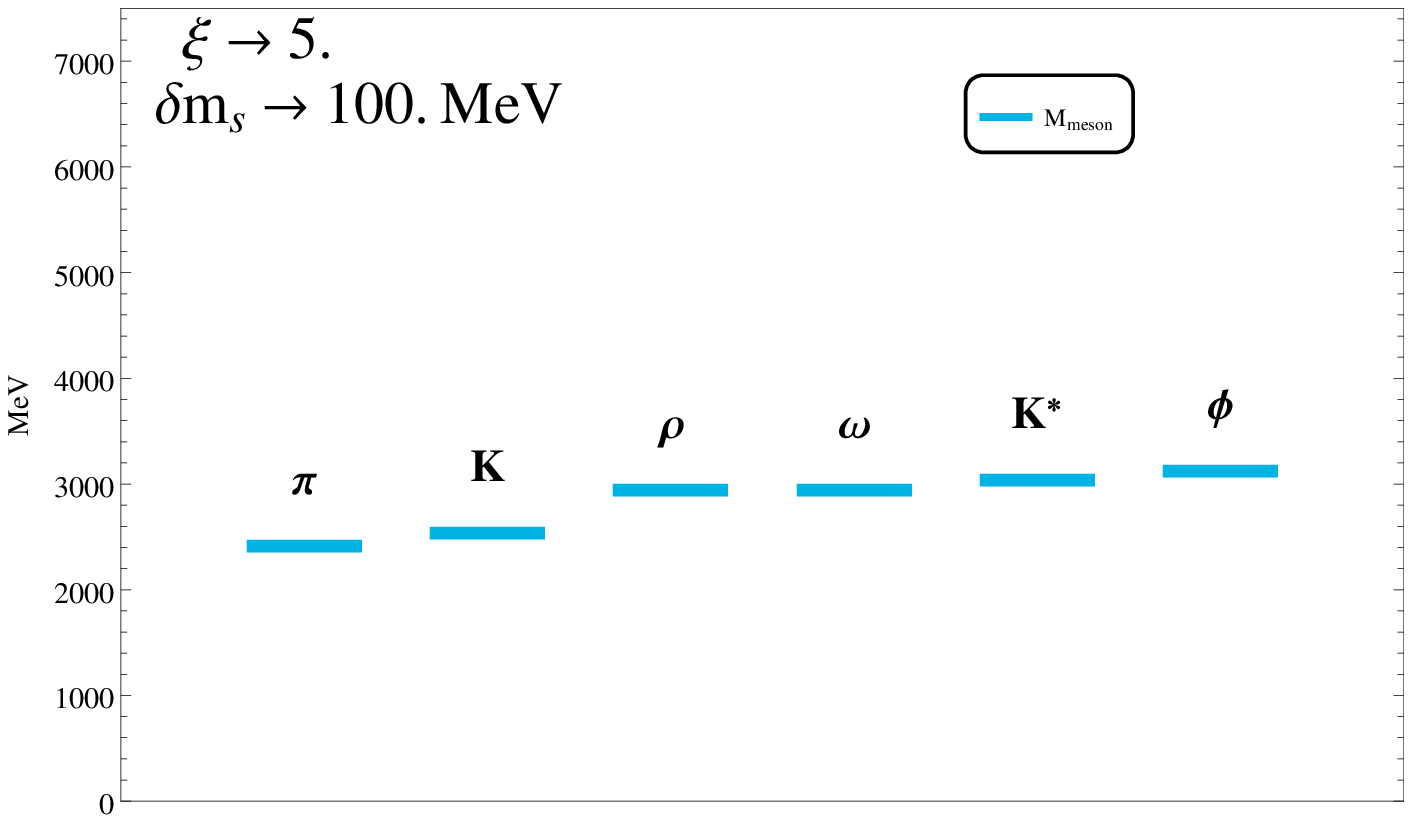} \\ 
    \includegraphics[width=.47\textwidth]{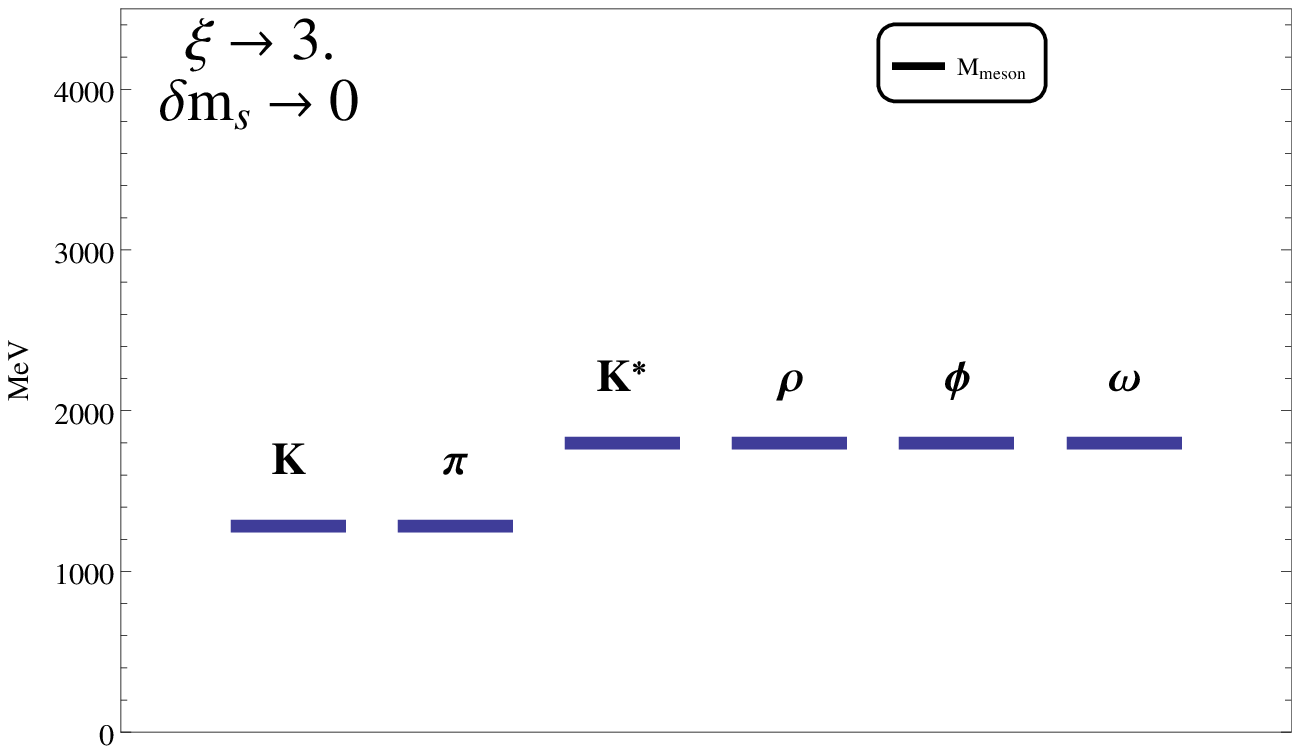}\quad
    \includegraphics[width=.47\textwidth]{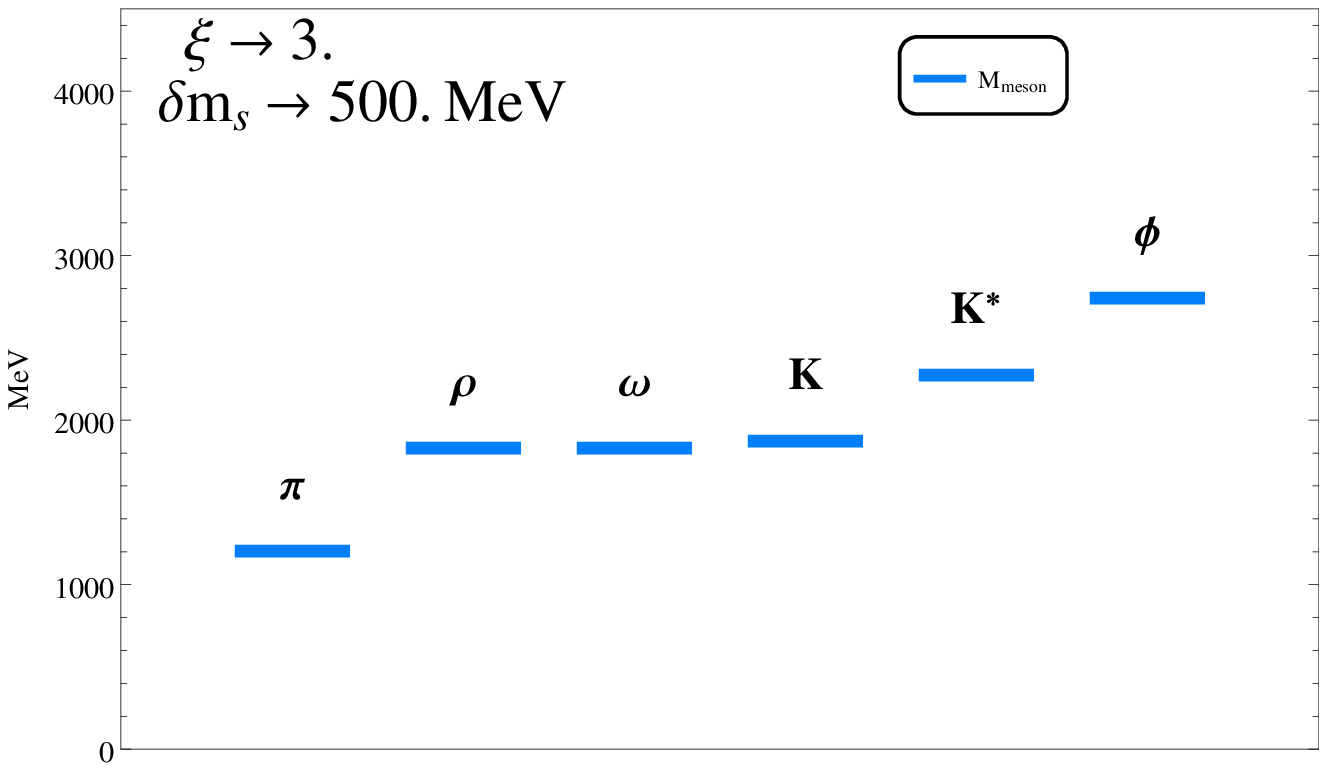}    
   \caption{Three flavour (uds) meson spectra for $\xi=3,5$ and semi-light state values given by $\delta m_s$.}
    \label{fig:mesondark1}
  \end{minipage}\\[1em]
  \end{figure}
   
\begin{figure}[h]
  \begin{minipage}{\textwidth}
    \centering
    \includegraphics[width=.47\textwidth]{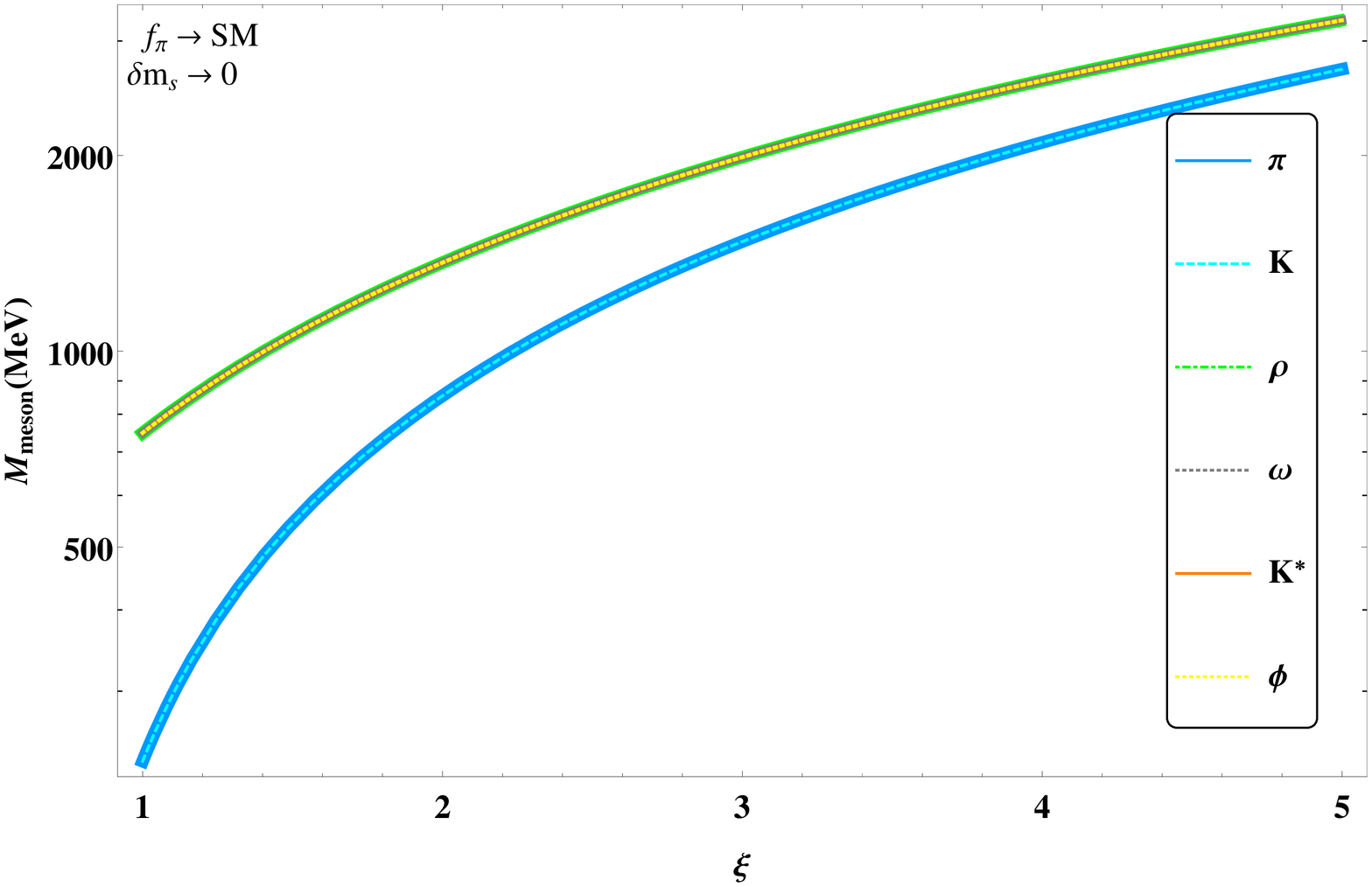}\quad
    \includegraphics[width=.47\textwidth]{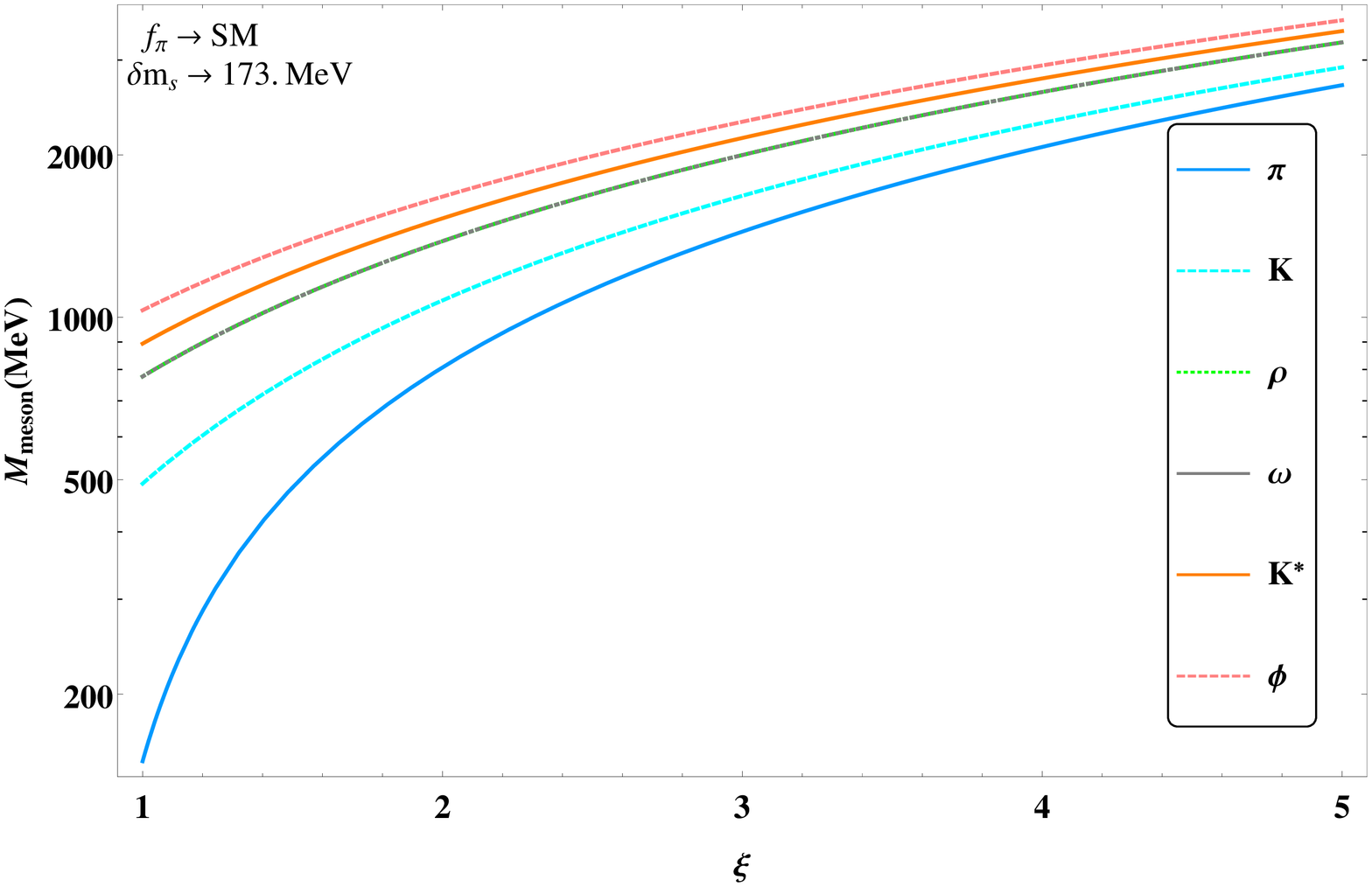} \\   
   \caption{Scaling of the meson spectra with $\xi$ for a given value of $\delta m_s$ shown in upper left. 
   The value of the pion decay constant, and thus the strength of chiral symmetry breaking is taken to remain constant.}
    \label{fig:mesonrun}
  \end{minipage}\\[1em]
  \end{figure}

Figure \ref{fig:mesondark1} shows the light meson spectra for a small set of different dark QCD cases. As the dark confinement scale is large, 
we take the anomalous meson to be sufficiently heavy that it is not part of the light set as discussed previously. In Figure \ref{fig:mesonrun} we examine how a sample of the meson spectra in this model varies with $\xi$. 
In particular we see the variation of the mass between the pseudoscalar and vector mesons as $\xi$ increases.
As we are assuming a consistent value of the pion decay constant
it must be the case that the bare quark masses are similarly increasing with $\xi$. This reiterates the previous statement that this model is not well suited to exploring the 
chiral limit and, indeed, exploring the full parameter space of varying the bare quark mass and confinement scale independently for a dark QCD is a task that chiral perturbation theory or lattice QCD studies may have the capacity to accomplish.

\section{Cosmological history of Dark Matter}\label{sec:NDM}
Asymmetric dark matter models position dark matter as the remaining abundance of matter in a dark sector 
following the annihilation of near equal amounts of dark matter and dark antimatter, similar to the baryon asymmetry in our own sector \cite{Petraki:2013wwa}.
We then require from an asymmetric dark matter model with a dark QCD a number of key features.
The first is that dark matter is stable and so a conserved global quantum number 
is necessary. Dark QCD can provide this in the form of a dark baryon number, however this 
must be present in the model of the dark quarks themselves, as is the case in models with a mirror symmetry.
Secondly, in order for the dark symmetric matter-antimatter components to be annihilated, 
a form of dark radiation such as dark electromagnetism U(1)$_D$ is needed or annihilation into ordinary radiation must occur. 
In the case that the symmetric component of dark matter annihilates into relativistic standard model species, the dark sector 
could avoid any constraints from the CMB data and nucleosynthesis by having a broken dark U(1) symmetry with a massive dark photon and therefore have zero dark relativistic species.
If dark photons fulfill that role of annihilating the symmetric component then this has direct implications as discussed in the previous chapter on the ordering of the hadronic spectra as well as the self-interaction cross section for dark matter.
The existence of a dark photon opens up the possibility of a kinetic mixing term $\frac{1}{2}F^{\mu \nu}F_{\mu \nu}'$. 
This can have important consequences for the interaction of DM with SM species, which has been explored in a number of asymmetric dark matter models such as \cite{Cline:2013zca, Cohen:2010kn}.
Additionally such light degrees of freedom in the dark sector contributes to $N_{eff}$ and will bring it above the current limits imposed if the temperature of the dark sector is equal to the visible sector. For this reason we consider in further detail
how the bounds on the number of effective neutrinos can still be compatible with such dark sectors.
If the contribution to $N_{eff}$ is suppressed by a lower temperature of the dark sector then one or more light species can fill this role in dark sector models. 
We will examine how such temperature differences can naturally be created in models with dark confinement transitions in the following section.
\subsection{Thermal history of dark sectors}
The total number of relativistic degrees of freedom is given by the equation,
\begin{equation}
\label{eq:degs}
 g_{*}(T)= \sum_b g_i \left(\frac{T_i}{T}\right)^4 + \frac{7}{8} \sum_f g_i \left(\frac{T_i}{T}\right)^4.
\end{equation}
The first term in equation \ref{eq:degs} is for bosonic and the second term is for fermionic degrees of freedom. We can see that
each relativistic species contributes to varying degrees depending on its relative temperature to our own sector.
One way of obtaining a lower temperature for a composite dark sector is to thermally decouple the two sectors at a moment in time with a large imbalance in the number of degrees of freedom. We consider the case that the 
two sectors are in thermal equilibrium, and thus at equal temperatures, from the early universe to a temperature $T_{\text{DEC}}$ after which the two temperatures, $T_V,T_D$, evolve independently. The relationship between the temperature of the two sectors
after thermal decoupling will depend on the number of degrees of freedom of each sector at the moment of decoupling and on the number of degrees of freedom of each sector at each moment in their evolution afterwards. 
By the conservation of entropy we have
\begin{equation}
\frac{T_V^3}{T_D^3}= \frac{g_D}{g_V} \left(\frac{g_V}{g_D}\right)_{\text{DEC}}.
\end{equation}
In particular, a lower temperature dark sector is possible if the visible sector has a greater number of degrees of freedom at the moment of decoupling. 
This is due to most of the entropy density being shifted to the visible sector, after which the two sectors evolve with the entropy density they had at the moment of $T_{\text{DEC}}$.
Models of composite dark matter, in which the dark composite scale is greater than the QCD scale, naturally come with such a moment in the thermal history.
At the confining transition in the standard model, we transition from $g_{*} \sim 61.75$ to  $g_{*} \sim 10.75$. 
This is due to all of the gluons and the three remaining relativistic quarks suddenly becoming non-relativistic as they form hadrons.
As noted in \cite{Farina:2015uea}, it is then possible to obtain a large temperature difference if decoupling takes place after the dark quark-hadron phase transition but before the quark-hadron phase transition of the visible sector.
This will depend however on the rest of the particle content of the dark sector. If we consider the case  of an otherwise mirror sector, we obtain a temperature ratio of $T_D/T_V \simeq 0.56$.
The distribution of light and heavy dark quarks is not critical here as the visible sector masses are known and
we are considering the region were the dark sector's degrees of freedom have lost all remaining dark colored particles. 
With this temperature difference, a mirror photon and three light mirror neutrino species we obtain $\Delta N_{eff}\simeq 0.73$ which is above the bounds.
However with the current bounds of $\Delta N_{eff}=0.11 \pm 0.23$, we see that dark sectors with fewer relativistic species are permissible.
With fewer mirror neutrino flavours we both increase the temperature difference and reduce the total contribution to $N_{eff}$. 
Assuming a standard model-like temperature difference between mirror photon and mirror neutrinos 
due to mirror electron injection we obtain $\Delta N_{eff}\simeq 0.49$ for two mirror neutrino flavours, 
$\Delta N_{eff}\simeq 0.31$ for one and $\Delta N_{eff}\simeq 0.17$ for none.
We can see that one light mirror neutrino and one mirror photon falls within the bound for $\Delta N_{eff}$ in this minimal case.
\footnote{Additional mirror neutrino flavours may be allowed in the case of more mass thresholds that could be crossed in the dark sector that would 
alter its total number of degrees of freedom in the temperature region between $\Lambda_{QCD}$ and $ \Lambda_{DM}$. One can also consider how models with a dark photon and a second relativistic fermion species, such as a dark neutrino, might 
have these relative dark temperatures be affected by additional injection mechanisms, similar to the electron annihilations in the standard model.}
The use of this mechanism requires a suitable coupling between the two sectors such that thermal decoupling will take place in the range $\Lambda_{QCD} < T_{\text{DEC}} < \Lambda_{DM}$.
This can  occur in one of two ways. Either the interaction rate falls below the Hubble rate in this temperature region or the number densities of the species involved become Boltzmann suppressed during this time.
In \cite{Farina:2015uea} an effective four fermion operator is considered and the effective cut-off scale is chosen such that the interaction rate meets the rate of expansion in the temperature range between the two quark-hadron phase transitions.

\subsection{Self-Interaction of dark baryons}
If the dark EM coupling is of the same scale as the SM value, the self interaction rate for a population of purely charged dark matter may be above the current bounds. For a recent analysis of constraints see 
\cite{Agrawal:2016quu} where DM with a U(1) gauge charge was considered and constraints from triaxiality and galaxy cluster mergers were compared to the significant bounds, 
in particular that for $\sim 10 \text{GeV}$ scale DM  the dark EM coupling $\alpha_D <10^{-4}$. Note that these bounds assume no compact object formation in the dark sector.
The strength of the dark gauge coupling in this case allows for different fractions of DM to be charged.
If the set of light quarks allows for neutral states however then we have seen that degenerate quark masses motivate models in which the neutral states are the lightest stable composites and if 
nuclear forces are  sufficiently weak then dark matter can satisfy these constraints.
In the case of a dark neutron analogue we consider the effect of having dark matter with no direct dark U(1) charge but rather a dark magnetic moment. 
In models that constraint the size of $\alpha_D$ for an unbroken U(1) symmetry of the dark sector, the effective drag force for charged particles interacting through 
a long range coulomb force can be modeled as a function of the Coulomb logarithm $\text{Log}(b_{\text{max}}/b_{\text{min}})$, with maximum and minimum impact parameters b.
In Ref. \cite{Ackerman:mha} the strongest constraint comes from galactic halo dynamics. By demanding that DM-DM interactions induce no more than a small fractional change
in the energy of a DM particle in a galactic halo on cosmological time scales we can obtain constraints on the size of $\alpha_D$, the dark sector's fine structure constant.
For the case of a population of magnetic dipoles, the radial force will scale as ($1/r^4$) and so we consider the effect of numerous long range collisions by 
deriving the effective drag that instead of scaling with the Coulomb Log, depends on the impact parameters through a term $\sim (1/b^4_{\text{min}}-1/b^4_{\text{max}})$.
The maximum impact parameter is commonly associated with the Debye length, however in the magnetic case we can see that the ($1/r^4$) suppression makes this less relevant, as 
even with no long range cut-off the magnetic term is no longer divergent, by contrast with the Coulomb interaction.
With a minimum impact parameter given by the de Broglie limit and an unconstrained maximum impact parameter, long range magnetic self-interactions permit $\alpha_D$ to be above the SM value in the region $ m_{\text{DM}} \geqslant 1 \text{GeV}$ while maintaining consistency with the phenomenological bounds.  
For short range interactions we can compare directly to the neutron-neutron scattering rate, $\sigma_{nn} \simeq 10^{-24} \, \text{cm}^{-2}$, which is just in the region of the self interacting dark matter constraints for this mass region. With $\xi$ increasing we expect this
cross section to decrease by $1/\xi^2$ as the physical size of dark baryons decreases. 
This shows that for a population dark neutrons, the favored dark matter candidate for most of the cases we have considered, all of the requirements of a dark matter candidate are satisfied.
In the absence of strong nuclear forces, these dark neutrons undergo no dissipation effectively and behave as a collisionless gas. 
Of course light mesons may accommodate long range nuclear forces and this will place additional limits on the meson spectra for a model of a composite dark sector.
In the case of dark glueball dark matter Ref. \cite{Cline:2013zca} considers the 
relationship derived from the suspected glueball state of QCD that $m_{GB} \sim 5.5 \Lambda_{DM}$ and estimates a self interaction cross section
$\sigma \sim 4 \pi / \Lambda_{DM}^2$.

\subsection{Dark sector abundances}
In order to find a natural explanation for the observed components of the universe, $\Omega_{D} \simeq 5 \Omega_{V}$, we require both a reason for the similar 
abundance of baryons and dark baryons as well as the similarity in mass. The relationship in the abundance can be formulated in a large number of different ways as 
it is dependent on the theory of the generation of the baryon asymmetry of our own sector if we are to regard the similarity as not a coincidence. 
In the case of mass, Section \ref{sec:HS} has demonstrated how the lightest stable 
baryon may scale with a dark confinement scale. 
\\\\
In the cases where there is a large mass gap between the lightest baryon and the rest of the spectra we can take the dark matter candidate to be this stable 
state if dark weak interactions exist and are not suppressed and dark quarks masses are light. 
If however the mass differences between two or more of the lightest dark QCD states are small then the dark QCD phase transition may produce similar numbers of these states.
This can be compared to standard cosmology where the near degeneracy of the neutron and proton produces roughly equal numbers following the quark hadron phase transition. 
In that case the near equal numbers allows for the process of nucleosynthesis where an array of stable
states of multiple nucleons can be formed. This is in the case where the dark sector has $n + \nu_e \leftrightarrow p + e^{-}$ interactions that 
maintain near equal n,p densities prior to the freeze out of weak interactions.
In models of mirror matter with dark electrons and neutrinos with masses larger than the mass difference of the lightest dark baryons such processes will be kinematically suppressed.
This equilibration can allow for dark helium to make up a significant fraction ($\sim 26 \%$) of the visible mass density.  
One can then consider dark sectors where two or more near degenerate composite states, that is where $\Delta m << \Lambda_{DM}$, 
are bound by dark nuclear forces into a complex arrangement of nuclei like objects. The complexity
may be far greater than standard nuclear theory where there is an approximately linear relationship between the number of protons and neutrons in nuclear bound states, 
for example; three near degenerate baryons, as in the case of degenerate u,d,s have six possible dibaryon states and
the mass hierarchy among these will depend non-trivially on the dark nuclear-like interactions. 
Compact objects in the form of neutron star-like bodies could also manifest ultimately depending on the model and the self interaction strength from strong-like interactions.

\section{Conclusion}\label{sec:Conclusion}
In this work we have considered the hypercentral approximation of the constituent quark model and the possible properties 
of a dark sector with a QCD analogue. In particular we have examined the dependence of the hadron spectra on the number of light chiral fermions and the
resulting phenomenology of dark QCD with a confinement scale in the few $\text{GeV}$ range.
As a class of theories to explain the nature of dark matter we have seen that larger confinement scales result 
in higher degrees of degeneracy in the spectra 
while the number of flavours has a significant impact on the mass and nature of the lightest baryon and meson for SU(3) theories. Constituent quark models have provided insight into the
nature of QCD and while many frameworks for advancing these calculations have sought to better replicate the
experimental signatures, potential models still allow us to probe the ground state spectra
in a simple manner with fewer parameters and a direct relationship to the confinement scale. By incorporating
such descriptions of dark QCD into asymmetric dark matter models
where the gauge couplings of the SM and hidden sectors are connected in the UV, the similarity in mass scale of 
DM and baryons finds a natural explanation. In this work we have found that if the hidden sector 
contains symmetries that parallel the visible sector, the higher confinement scale motivates theories with
a neutral ground state in addition to higher degeneracy among the baryons with different total spin and less degeneracy in charge. The spin and charge of the ground state is further
dependent on the number of light dark quarks in the theory and the quantum numbers of these quarks that make up the light set.
Given the unexplained almost five orders of magnitude that the quark flavours of the SM span, the possibility of a 
non-trivial splitting arrangement in the set of dark quarks allows for the possibility
of a very wide variety of composite dark matter models that differ greatly from conventional chromodynamics.

\section*{Appendix}
\setcounter{equation}{0}
\renewcommand{\theequation}{A\arabic{equation}}
The full potential including additional spin-spin interactions, isospin-isospin and spin-isospin interactions has the form
\begin{equation}
\label{eq:potential incl ss}
H_{V}= V(\vec{r_{ij}}) + V_{SS}(\vec{r_{ij}}) \, \vec{\sigma}_1 \cdot \vec{\sigma}_2 + V_{II}\,(\vec{r_{ij}}) \vec{t}_1 \cdot \vec{t}_2 + V_{SI}\, (\vec{r_{ij}}) (\vec{t}_1 \cdot \vec{t}_2)(\vec{\sigma}_1 \cdot \vec{\sigma}_2)
\end{equation}	
and the full non-relativistic Hamiltonian is then \cite{Isgur:1978xj, Isgur:1977ef} 
\begin{equation}
\label{eq:Hamiltonian + HI}
H = \sum_{i}m_i+ H_0 + H_{V}
\end{equation}
with $H_0 = \sum_{i} \frac{p_i^2}{2m_i}$. As the spin-spin interaction has a larger contribution to the potential than the remaining hyperfine interactions and the spin-orbit term is taken to be negligible as in \cite{Giannini:2012vy} we similarly 
focus on a model with the spin-spin effect contributing the most important effects to mass differences. It has the form \cite{DeRujula:1975qlm}
\begin{equation} \label{eq:HIham}
H^{ij}_{HI} = A\left[(\frac{8\pi}{3}) \, \vec{S_i} \cdot \vec{S_j} \, \delta^3(r_{ij}) \right],
\end{equation} 
where $A = \frac{2 \alpha_s}{3m_im_j}$. In the case of the confining potentials of Eq. \ref{eq:potentialvariation} where analytic solutions are not obtainable we use the matrix methods of \cite{Richard:1992uk}.
This then uses the expansion of the 6D hyperspherical Schrodinger equation using the Fourier expansion of the spatial wavefunctions over the hyperradius x. Following the matrix methods converts this to a scaled coordinate 
$y=\frac{x}{x+r_0}$ where $r_0$ remains as a scaling estimate of the radius of the spatial wavefunction. We can then express the hypercentral wavefunctions as
\begin{equation}
\psi(y)=\sum_{i=1}^N a_i \, \sin(i\pi y ).
\end{equation}
This reduces the differential equation to a matrix eigenvalue problem that gives the first N levels for a given value of $\gamma$, 
\begin{flalign}
&\sum_{j} \left[\left[ \frac{1}{2 m} \frac{(1-y_i)^4}{r_0^2} \sum_k (\frac{2}{N+1}) \sin(k \pi y_j)\, k^2 \pi^2 sin(k \pi y_i)\right] \right. \\ \nonumber
&- \left[\left.  \frac{1}{2m}\frac{5}{r_0^2} \frac{(1-y_i)^3}{y_i} \sum_c \frac{2}{N+1} \sin(c \pi y_j)\,  c \pi \cos(c \pi y_i)\right] \right.\\
&+ \left( \frac{1}{2m}\frac{\gamma(\gamma+4)}{x(y_j)^2}+ V(x(y_j))\right)\delta_{ij}  \biggr] \psi_j= E_{N [\gamma]} \psi(y_i).    \nonumber
\end{flalign}
This can be compared to the numerical solution of the case without the hypercentral approximation. In that calculation, a similar change of variables allows for the calculation of the complete set of coupled hyperspherical differential equations. 
In our case we apply it to a variety of non-analytic potentials that scale to dark sector parameters and that use the hypercentral approach.


\bibliography{DQCD.bib}


\end{document}